# Crack tip fields and fracture resistance parameters based on strain gradient plasticity

V. Shlyannikov[1], E. Martínez-Pañeda[2], A. Tumanov[1] and A. Tartygasheva[1]

[1]*Institute of Power Engineering and Advanced Technologies*
*FRC Kazan Scientific Center, Russian Academy of Sciences, Kazan, Russian Federation*
[2]*Department of Civil and Environmental Engineering, Imperial College London, London SW7 2AZ, UK*

**Abstract.** The crack tip mechanics of strain gradient plasticity solids is investigated analytically and numerically. A first-order mechanism-based strain gradient (MSG) plasticity theory based on Taylor's dislocation model is adopted and implemented in the commercial finite element package ANSYS by means of a user subroutine. Two boundary value problems are considered, a single edge tension specimen and a biaxially loaded plate. First, crack tip fields are characterized. Strain gradient effects associated with dislocation hardening mechanisms elevate crack tip stresses relative to conventional plasticity. A parametric study is conducted and differences with conventional plasticity predictions are quantified. Moreover, the asymptotic nature of the crack tip solution is investigated. The numerical results reveal that the singularity order predicted by the first-order MSG theory is equal or higher to that of linear elastic solids. Also, the crack tip field appears not to have a separable solution. Moreover, contrarily to what has been shown in the higher order version of MSG plasticity, the singularity order exhibits sensitivity to the plastic material properties. Secondly, analytical and numerical approaches are employed to formulate novel amplitude factors for strain gradient plasticity. A generalized *J*-integral is derived and used to characterize a nonlinear amplitude factor. A closed-form equation for the analytical stress intensity factor is obtained. Amplitude factors are also derived by decomposing the numerical solution for the crack tip stress field. Nonlinear amplitude factor solutions are determined across a wide range of values for the material length scale *l* and the strain hardening exponent *N*. The domains of strain gradient relevance are identified, setting the basis for the application of first-order MSG plasticity for fracture and damage assessment.

1. ## Introduction

Strain gradient plasticity models have enjoyed significant attention in the past 20 years. Constitutive theories accounting for the role of plastic strain gradients, and their associated length scale parameters, have enabled capturing the size effects observed in metals at small scales as well as regularizing otherwise ill-posed boundary value problems at the onset of material softening (Engelen et al., 2006; Voyiadjis and Song, 2019). The need for constitutive relations involving plastic strain gradients has been motivated by micro-scale experiments such as wire torsion (Fleck et al., 1994; Guo et al., 2017), or bending and constrained shear of thin foils (Stolken and Evans, 1998; Mu et al. 2014). These and other experiments have revealed a notable size effect in metals, with the effective flow strength increasing 3-fold by reducing specimen size (*smaller is stronger*). This strengthening and hardening effects, not observed in uniaxial tension samples of equal dimensions, are attributed to the presence of geometrically necessary dislocations (GNDs), which are induced by non-homogeneous plastic deformation (Ashby, 1970). The observed elevation of the flow stress with diminishing size can be quantitatively captured using strain gradient plasticity models, which extend conventional J2 flow theory of plasticity by incorporating a dependence on gradients of plastic strain (Fleck and Hutchinson, 1997; Gurtin and Anand, 2005). To enable dimensional matching, a characteristic material length scale parameter $l$ is introduced, which characterizes the capacity of the material to harden due to the generation of GNDs.

Local strengthening due to plastic strain gradients can also be observed in large (e.g., macro-scale) samples when plastic deformation is heterogeneous and is confined to a small region, as near the edge of an indenter (Poole et al. 1996) or at a crack tip (Wei and Hutchinson, 1997; Martínez-Pañeda and Niordson, 2016). In both stationary and propagating cracks, the use of strain gradient plasticity reveals a notable stress elevation in the vicinity of the crack tip, relative to conventional plasticity predictions (Wei and Qiu, 2004; Komaragiri et al., 2008; Mikkelsen and Goutianos, 2009; Seiler *et al.*, 2016; Kristensen *et al.*, 2020). This stress elevation is grounded on local strengthening due to the role of strain gradients and the associated large dislocation densities that originate near the tip of cracks. The high crack tip stresses predicted by strain gradient models has provided a mechanistic rationale to interpret a number of damage mechanisms, such as atomic decohesion in the presence of plasticity (Qu *et al.*, 2004; Fuentes-Alonso and Martínez-Pañeda, 2020), fatigue (Sevillano, 2001; Pribe et al., 2019), cleavage fracture of ferritic steels (Qian et al, 2011; Martínez-Pañeda et al., 2019) or hydrogen embrittlement (Martínez-Pañeda *et al.*, 2016; Fernández-Sousa *et al.*, 2020). The aim has been to link scales in fracture mechanics by enriching continuum theories to properly characterize material behavior at the small scales involved in crack tip deformation (Hutchinson, 1997).

Attention has also been focused on understanding and characterizing the asymptotic nature of the crack tip stress distribution predicted by the various classes of strain gradient plasticity models. For example, Xia and Hutchinson (1996) and Huang *et al.* (1995,1997) examined the crack tip singular behavior of solids characterized by coupled-stress theories without stretch gradients. Chen *et al.* (1999) investigated the crack tip asymptotic behavior of a gradient model incorporating the gradients of the elastic strains. These analyses follow a classic HRR (Hutchinson, 1968; Rice and Rosengren, 1968) approach, and determine the asymptotic nature of the stress solution by neglecting elastic contributions. More recently, Martínez-Pañeda and Fleck (2019) showed, analytically and numerically, that phenomenological higher-order theories (Gudmundson, 2004; Fleck and Willis, 2009) exhibit an elastic region adjacent to the crack tip, implying that the solution cannot be simplified by ignoring the elastic contribution. Alternative formulations to these phenomenological models are the so-called mechanism-based strain gradient (MSG) plasticity theories, which are grounded on Taylor's dislocation model (Taylor, 1938; Nix and Gao, 1998; Gao *et al.*, 1999). Both higher order and lower order formulations have been proposed, the latter often referred to as conventional mechanism-based strain gradient (CMSG) plasticity (Huang et al., 2004). Jiang et al. (2001) and Shi et al. (2001) investigated the asymptotic behavior predicted by MSG plasticity, revealing a stress singularity larger than that predicted with linear elasticity.

In the present work, analytical and numerical methodologies are used to characterize the crack tip behavior of CMSG plasticity solids. The finite element simulations conducted reveal notable differences with its higher order counterpart; the crack tip field appears not to have a separable solution and the singularity order is sensitive to the plastic properties of the material. Also, a generalized J-integral is defined and used to derive amplitude stress intensity factors for the first time. A plastic stress intensity factor (Hilton and Hutchinson, 1971) is defined for strain gradient plasticity, using both numerical and analytical approaches. The potential applications of these fracture resistance parameters are discussed.

## 2. Constitutive equations of conventional mechanism-based strain gradient (CMSG) plasticity theory

The conventional theory of mechanism-based strain gradient plasticity (CMSG) developed by Huang et al. (2004) has enjoyed great popularity due to its simpler numerical implementation, relative to higher order models. The strain gradient effect comes into play via the incremental plastic modulus not requiring the consideration of higher order stresses and higher order boundary conditions. Thus, it can be implemented numerically as a user material using standard finite element formulations. Both MSG plasticity (higher order) and CMSG plasticity (lower order) are based on the same theoretical framework: Taylor (1938) dislocation model. Thus, the shear flow stress $\tau$ is defined as a function of the shear modulus $\mu$, the Burgers vector $b$ and the dislocation density $\rho$ as:

$$\tau = \alpha \mu b \sqrt{\rho}, \tag{1}$$

where $\alpha$ is an empirical coefficient ranging from 0.3 to 0.5. The dislocation density $\rho$ is composed of the density $\rho_S$ for statistically stored dislocations (SSDs), which accumulate by trapping each other in a random manner, and density $\rho_G$ for geometrically necessary dislocations (GNDs), which are required for compatible deformation of various parts of the material, i.e.,

$$\rho = \rho_S + \rho_G. \tag{2}$$

The SSD density is related to the flow stress and the material stress-strain curve in uniaxial tension

$$\rho_S = \left[\sigma_{ref} f\left(\varepsilon^P\right) / M \alpha \mu b\right]^2 \tag{3}$$

The GND density is related to the curvature of plastic deformation, or the effective plastic strain gradient $\eta^P$, by

$$\rho_G = \bar{r}\frac{\eta^P}{b}, \tag{4}$$

where $\bar{r}$ is the Nye factor, which is around 1.90 for face-centered-cubic polycrystals. The measure of the effective plastic strain gradient $\eta^P$ was reported by Gao et al. (1999) in the form of three quadratic invariants of the plastic strain gradient tensor $\eta^p_{ijk}$ as:

$$\eta^p = \left(c_1 \eta^p_{iik}\eta^p_{jjk} + c_2 \eta^p_{ijk}\eta^p_{ijk} + c_3 \eta^p_{ijk}\eta^p_{ijk}\right)^{1/2}. \tag{5}$$

The coefficients were determined to be equal to $c_1 = 0$, $c_2 = 1/4$ and $c_3 = 0$ from three dislocation models for bending, torsion and void growth. The components of the plastic strain gradient tensor $\eta^p_{ijk}$ are computed from the plastic strain tensor $\varepsilon^p_{ij}$ as:

$$\eta^p_{ijk} = \varepsilon^p_{ik,j} + \varepsilon^p_{jk,i} - \varepsilon^p_{ij,k} \tag{6}$$

The tensile flow stress is related to the shear stress by

$$\sigma_{flow} = M\tau = M\alpha\mu b\sqrt{\rho_S + \bar{r}\frac{\eta^P}{b}}, \tag{7}$$

where $M$ is the Taylor factor, taken to be 3.06 for fcc metals. Because the plastic strain gradient $\eta^P$ vanishes in uniaxial tension, the density $\rho_S$ for SSDs is described by Eq. (3), and the flow stress becomes

$$\sigma_{flow} = \sigma_{ref}\sqrt{f^2(\varepsilon^P) + l\eta^P}, \tag{8}$$

where

$$l = 18\alpha^2\left(\frac{\mu}{\sigma_{ref}}\right)^2 b \tag{9}$$

is the intrinsic material length in the strain gradient plasticity based on parameters of elasticity (shear modulus $\mu$), plasticity (reference stress $\sigma_{ref}$), and atomic spacing (Burgers vector $b$). For metallic materials, the internal material length is indeed on the order of microns, consistent with the estimate by Fleck and Hutchinson (1997). One should note that if the characteristic length of plastic deformation is much larger than the intrinsic material length $l$, the GNDs-related term $l\eta^p$ becomes negligible, such that the flow stress degenerates to $\sigma_{ref} f(\varepsilon^p)$, as in conventional plasticity. Also, we emphasize that although the intrinsic material length $l$ depends on the choice of $\sigma_{ref}$ via Eq. (9), the flow stress in Eq. (8) is independent of $\sigma_{ref}$, as both terms inside of the square root are independent of $\sigma_{ref}$. A potential choice for the reference stress and the material function $f(\varepsilon^p)$ is

$$\sigma = \sigma_{ref} f(\varepsilon^p) = \sigma_y \left(\frac{E}{\sigma_y}\right)^N \left(\varepsilon^p + \frac{\sigma_y}{E}\right)^N, \tag{10}$$

such that,

$$\sigma_{ref} = \sigma_y (E/\sigma_y)^N, \tag{11}$$

and the nondimensional function of plastic strain $f(\varepsilon^p)$ is determined from the uniaxial stress-strain curve, which for most ductile materials can be written as a power law relation

$$f(\varepsilon^p) = \left(\varepsilon^p + (\sigma_y/E)\right)^N. \tag{12}$$

In Eqs. (10)–(12), $\sigma_y$ denotes the initial yield stress, and $N$ is the plastic work hardening exponent ($0 \leq N < 1$).

The constitutive formulation described so far is common to both MSG and CMSG plasticity theories. Suitable constitutive choices were established by Huang *et al.* (2004) to circumvent the need to use higher order stresses work-conjugate to plastic strain gradients. Namely, a viscoplastic formulation was presented, in the form of the following constitutive equations

$$\dot{\varepsilon}^p = \dot{\varepsilon} \left(\frac{\sigma_e}{\sigma_{flow}}\right)^m = \dot{\varepsilon} \left[\frac{\sigma_e}{\sigma_{ref}\sqrt{f^2(\varepsilon^p) + l\eta^p}}\right]^m \tag{13}$$

$$\dot{\sigma}_{ij} = K\dot{\varepsilon}_{kk}\delta_{ij} + 2\mu\left[\dot{\varepsilon}'_{ij} - \frac{3\dot{\varepsilon}}{2\sigma_e}\left(\frac{\sigma_e}{\sigma_{flow}}\right)^m \sigma'_{ij}\right], \tag{14}$$

where $\sigma_e$ is the effective stress, $\dot{\varepsilon}'_{ij}$ is the deviatoric strain rate, and $m$ is the rate-sensitivity exponent. Values of $m$ larger than 5 have shown to provide a response very close to the rate-independent limit. In this work, $m = 20$ is adopted so as to simulate rate-independent behavior. Huang *et al.* (2004) compared CMSG plasticity with its higher order counterpart, MSG plasticity. The stress distributions predicted by the lower and higher order theories are only different within a thin boundary layer, whose thickness is approximately 10 nm. CMSG plasticity is a continuum theory and is therefore bounded at the lower end (e.g., it cannot be applied down to the nanometer scale). As it represents a collective behavior of dislocations, the lower limit of physical validity should be at least several times the dislocation spacing; i.e., larger than 10 nm. However, the fracture analyses of this work reveal notable differences with MSG plasticity beyond 10 nm ahead of the crack tip.

## 3. Boundary value problems and finite element implementation

Numerical solutions are obtained for two boundary value problems of particular interest, see Fig. 1: a single edge tension (SET) specimen and a biaxially loaded plate (BLP). The SET configuration is a well-known benchmark that has been widely used for the investigation of crack tip fields in strain gradient solids. The load is prescribed by imposing a displacement on the pins. We model the contact between the pins and the specimen by using a surface to surface contact algorithm with finite sliding (Fig.1a). The choice of the BLP configuration is grounded on the well-known fact that a plate with equibiaxial tension has a vanishing *T*-stress (Betegón and Hancock, 1991; Shlyannikov, 2013). This enables a direct comparison with the analytical results of the HRR theory when $l = 0$. The panel is subjected to two perpendicular and equal loads: one is parallel to the Y-axis, and the other one is parallel to the X-axis, i.e., forming an equibiaxial tension nominal stress state (Fig. 1b). The initial crack is located at the center of the biaxially loaded panel.

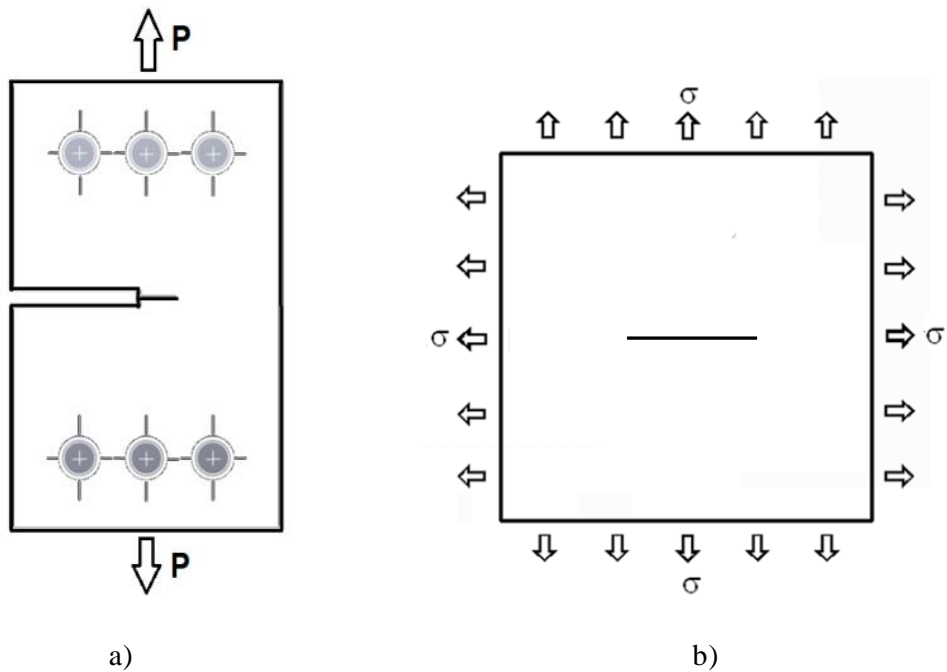

a)          b)

**Fig. 1.** Boundary value problems under consideration: (a) Single edge tension (SET) specimen, and (b) biaxially loaded plate.

One of the key features of our study is the evaluation of coupling material properties and strain gradient plasticity effects. To this end, a wide range of values for the plastic work hardening exponent $N$ and for the intrinsic material length parameter $l$ are used in our numerical calculations. Namely, $N$ is varied between 0.1 and 0.5 while the material length scale is varied between 1 and 10 μm. Changing the magnitude of the strain hardening exponent enables us to compare the resulting trends with those of the classic analytical HRR solution (Hutchinson, 1968, Rice and Rosengren, 1968) for conventional elastic-plastic solids. Regarding the length scale parameter, we aim at spanning the range of experimentally reported length scales, $l = 1 - 10$ μm (Fuentes-Alonso and Martínez-Pañeda, 2020), and we also consider the conventional plasticity case ($l = 0$). To compare results from the two different geometries, calculations will be presented as a function of the remote stress intensity factor characterizing the stress state, $K_I$. The analysis will be restricted to pure mode I conditions and the mode I stress intensity factor will often be normalized as $\bar{K}_1 = K_1 / \sigma_y \sqrt{l}$. The sensitivity of the results to the remote load will be explored and the quantity $K_1 / \sigma_y \sqrt{l}$ will be varied across a wide range; from 6.17 to 68.38. Stress quantities will be presented normalized by the initial yield stress $\sigma_Y$ and the distance to the crack tip is normalized by the characteristic material length: $r/l$. The crack faces remain traction-free in the two boundary value problems considered. The elastic stress intensity factor, $K_I$, of the remotely applied field increases monotonically, such that there is no unloading.

The theoretical framework described in Section 2 is implemented in the commercial finite element package ANSYS. This is achieved by using a user material subroutine USEMAT. Recall that, unlike the higher-order theory of mechanism-based strain gradient plasticity (Gao et al., 1999; Huang et al., 2000), CMSG is a first-order theory, not involving higher order stresses and with the same governing equations as conventional plasticity. The effect of the plastic strain gradients is taken into account by modifying the tangent modulus. Thus, a non-local formulation is used by which the computation of the plastic strain rate via (13) requires estimating the effective plastic strain gradient from the plastic strain components at the different integration points within the element. Fortran modules are used to store the plastic strain components across the Gaussian integration points, and the plastic strain gradient is computed by numerical differentiation within the element. This is accomplished by interpolating the plastic strain increment within each element via the values at Gaussian integration points in the isoparametric space, and subsequently determining the gradient of the plastic strain increment via differentiation of the shape function. The Newton-Raphson method is used to solve the resulting non-linear problem, similar convergence rates to those obtained using ANSYS's in-built plasticity models are observed. To the best of the authors' knowledge, the present work constitutes the first implementation of a mechanism-based strain gradient plasticity theory in ANSYS.

We model a SET specimen (Fig.1a) of width W = 35 mm and initial crack length $a_0$ = 14 mm. For comparison with the HRR solution for an infinite-sized cracked body, the BLP has a width W = 1000 mm and an initial crack length $a_0$ = 10 mm. In the finite element models, an initial crack tip is defined as an ideally sharp crack (mathematical notch). With the aim of accurately characterizing the influence of the plastic strain gradients, a highly refined mesh is used near the crack tip – see Fig. 2. A mesh sensitivity analysis is conducting, revealing that a characteristic element size $h$ below 16 nm delivers mesh-independent results. Representative results of the mesh sensitivity analysis are shown in Fig. 2. A characteristic element size of $h$=5 nm is chosen, ensuring mesh-independent results.

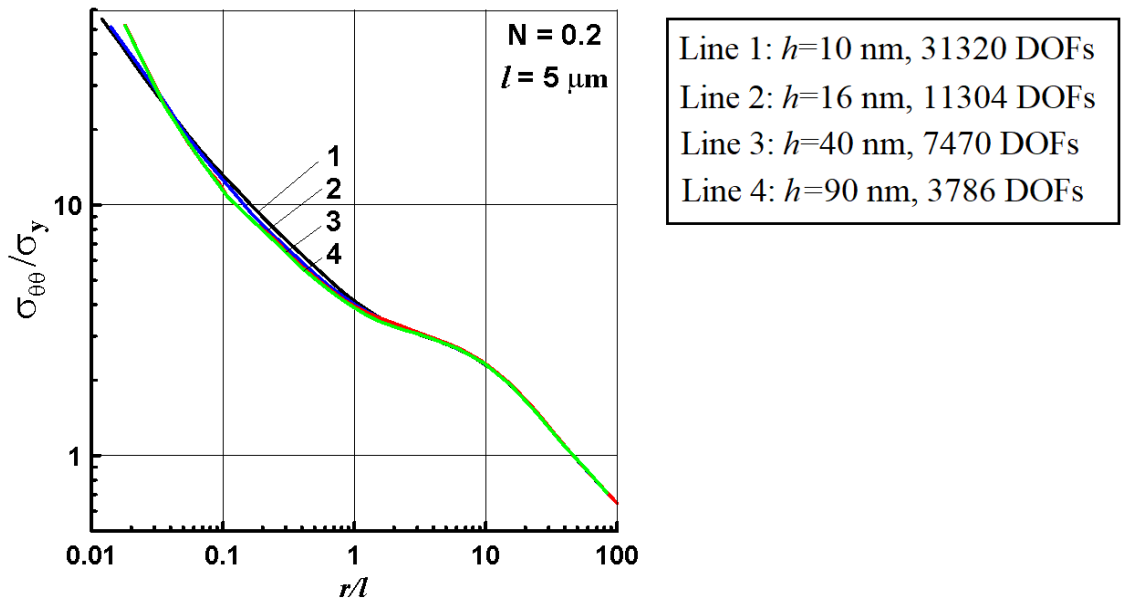

**Fig. 2.** Hoop stress distribution ahead of the crack: mesh-sensitivity analysis. Each line corresponds to a different model with characteristic element size $h$; the total number of degrees-of-freedom (DOFs) is also reported for each model.

Efforts were made to ensure an aspect ratio of one in the elements close to the crack tip. Quadrilateral, quadratic plane strain elements were used, the typical number being approximately 550,000. The numerical implementation was validated by reproducing the results of Qu et al. (2004) and Martínez-Pañeda and Betegón (2015).

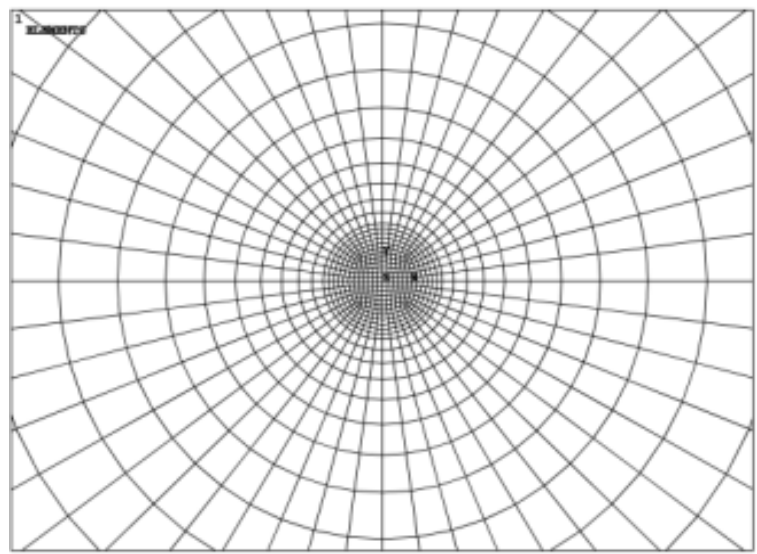

**Fig. 3.** Detail of the finite element mesh used in the vicinity of the crack tip. The characteristic element size is on the order of 5 nm.

## 4. Crack tip fields

We begin our analysis by providing a finite element characterization of crack tip fields. First, attention is drawn to the crack tip asymptotic behavior of CMSG plasticity solids. While the asymptotic behavior of solids characterized by phenomenological strain gradient theories and by the higher order MSG plasticity model is known (Shi et al., 2001; Martínez-Pañeda and Fleck, 2019), the investigation of crack tip asymptotics in CMSG plasticity has not been carried out yet. Finite element results for the effective von Mises stress distribution $\sigma_e$ ahead of the crack tip are shown in Fig. 4. Crack tip stresses are shown for both CMSG plasticity and conventional plasticity for a wide range of strain hardening exponent $N$ values. The results corresponding to the conventional plasticity solution are denoted as CPS. Numerical predictions for the single edge tension (SET) sample are shown in Fig. 4(a) while the results for the biaxially loaded plate (BLP) are shown in Fig. 4(b).

In both the SET and BLP cases, the remote stress equals $\sigma^\infty/\sigma_Y = 0.155$, which corresponds to a normalized stress intensity factor of $\bar{K}_1 = 30.58$ and $\bar{K}_1 = 14.54$, respectively. The material properties are $\sigma_Y/E = 0.1\%$, $\nu = 0.3$ and $l = 5$ μm. Outside of the plastic zone ($\sigma_e/\sigma_Y \leq 1$) the $r^{-1/2}$ slope inherent to linear elastic singularity is observed in all cases. For a given $N$ value and an effective stress magnitude slightly higher than the yield stress, the result follows the classic HRR solution for both CMSG and conventional plasticity (CPS).

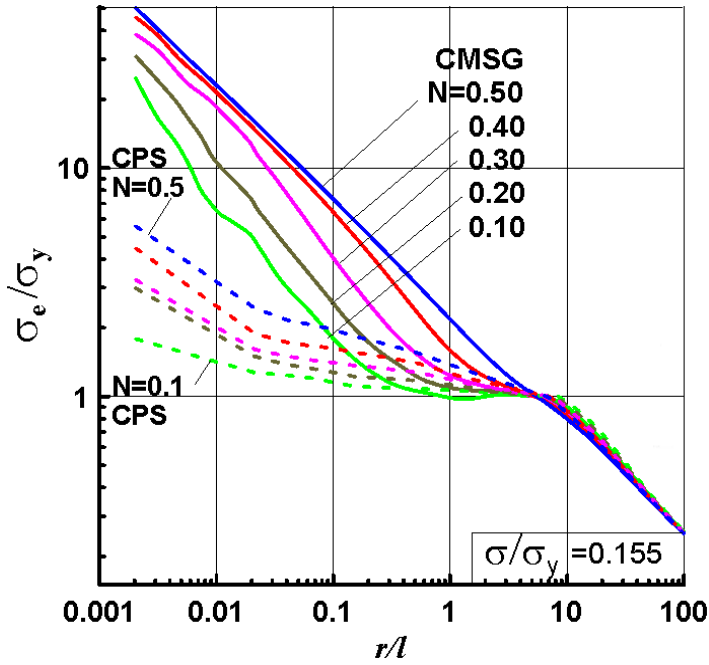

(a)

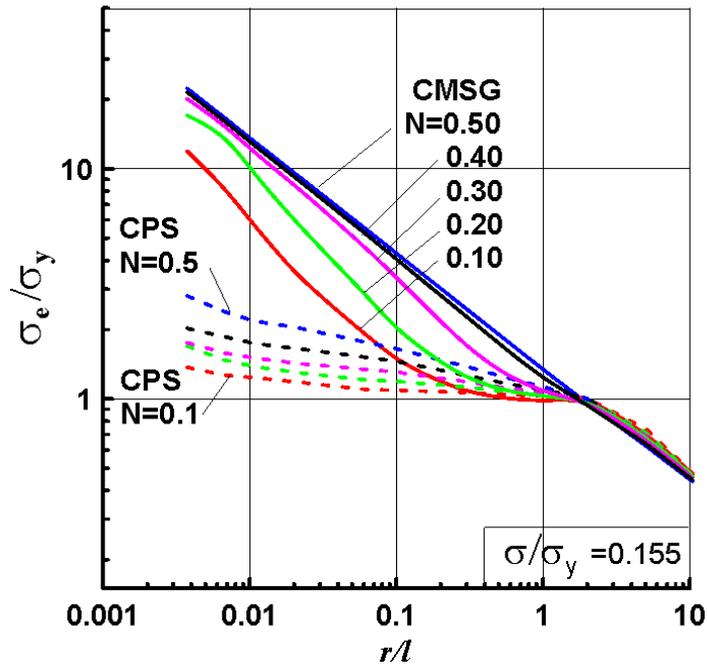

(b)

**Fig. 4.** Effective stress distribution $\sigma_e$ ahead of the crack tip for CMSG plasticity and conventional von Mises plasticity as a function of the strain hardening exponent: (a) SET sample for $\bar{K}_1 = 30.58$ ($\sigma^\infty/\sigma_Y = 0.155$) and (b) BLP sample for $\bar{K}_1 = 14.54$ ($\sigma^\infty/\sigma_Y = 0.155$). For N<0.5, the stress singularity exceeds that of elastic solids. Material properties: $\sigma_Y/E = 0.001$, $\nu = 0.3$ and $l = 5\ \mu m$.

However, when as $r$ gets smaller the effect of the plastic strain gradients becomes apparent and notable differences can be observed with the conventional plasticity predictions. Plastic strain gradients significantly elevate the stresses close to the crack tip, with $\sigma_e$ being up to an order of magnitude larger than the conventional plasticity solution. Strain gradient effects become important within a distance of $0.1l$ to $l$ of the crack tip, depending on the strain hardening exponent. This is in agreement with Xia and Hutchinson's (1996) estimate of the size of the gradient dominated zone and with Jiang et al. (2001) numerical calculations for MSG plasticity. Interestingly, the results shown in Fig. 4 for the CMSG plasticity predictions show sensitivity of the asymptotic stress singularity to the strain hardening exponent $N$. The stress distribution within a small distance to the crack tip $r/l \leq 0.1$ exhibits a slope higher than that of linear elastic solids, except for very high values of $N$ (>0.4) where the solution resembles that of linear elasticity. This behavior is observed for both the SET and BLP cases and has important implications. First, a stress singularity larger than the classic linear elastic result $r^{-1/2}$ entails a different asymptotic behavior relative to other classes of strain gradient models. As shown by Martínez-Pañeda and Fleck, (2019), phenomenological higher order strain gradient plasticity models such as those developed by Gudmundson (2004), Gurtin and Anand (2005) and Fleck and Willis (2009) exhibit an inner elastic $K$ field such that the stresses scale as $r^{-1/2}$ close to the crack tip. Shi et al. (2001) also found a higher stress singularity than linear elasticity in the case of the higher-order MSG plasticity model; they combined numerical and analytical techniques and concluded that the stresses scaled as $r^{-2/3}$ in the vicinity of the crack tip. This was used by Martínez-Pañeda et al. (2017) to develop enriched numerical schemes based on the extended finite element method. Thus, strain gradient plasticity models based on Taylor's dislocation theory appear to predict a stronger singularity than linear elastic solids, for both lower order and higher order formulations. Secondly, the stress singularity appears to be dependent on the strain hardening exponent, as it is the case of the HRR solution. This result is in disagreement with the numerical analyses by Jiang et al. (2001) and Shi et al. (2001), and would indicate a greater role of higher order stresses than previously anticipated.

The influence of the remote load, the yield stress and the material length parameter are assessed in Fig. 5. Dimensional analysis shows that the non-dimensional set governing the role of plastic strain gradients is $l/R_0$, where $R_0$ is the plastic zone size as defined by Irwin's approximation:

$$R_0 = \frac{1}{3\pi}\left(\frac{K_I}{\sigma_Y}\right)^2 \tag{15}$$

First, in Fig. 5a, crack tip stress distributions are shown for a fixed value of $l/R_0$ but two selected values of the normalized remote load $K_I/\sigma_y\sqrt{l}$: 3.2 and 14.7. Also, the strain hardening exponent is varied between 0.1 and 0.5. In agreement with expectations, the overall stress level increases with the remote load.

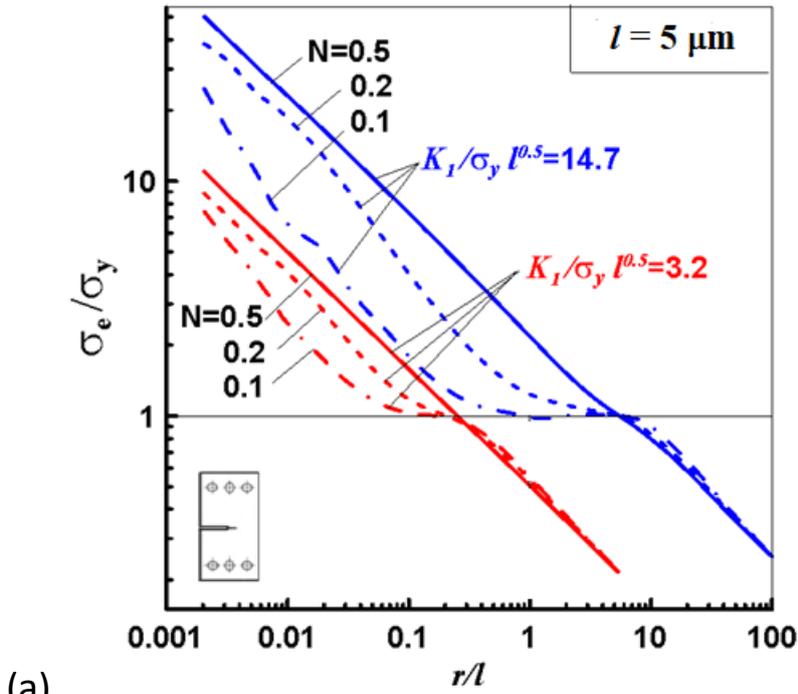

(a)

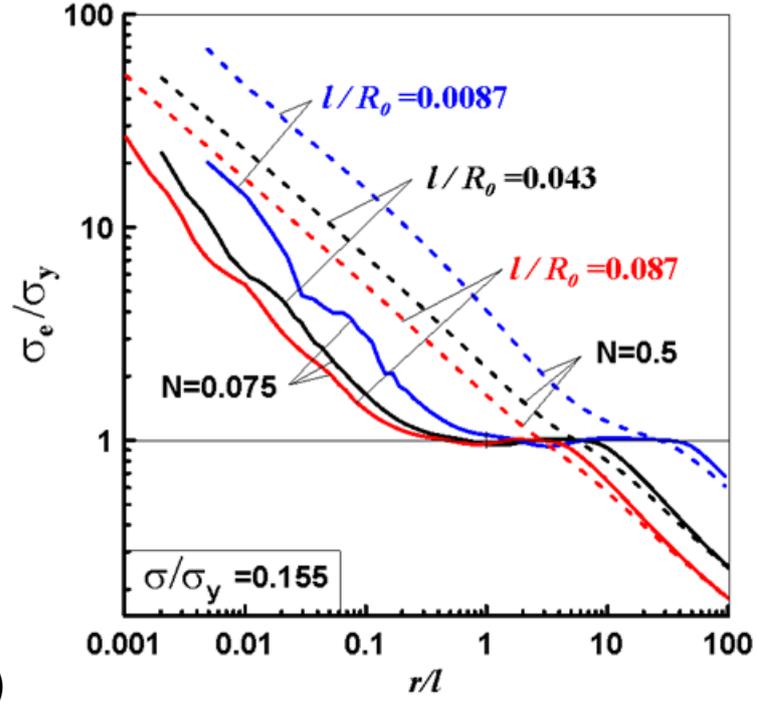

(b)

**Fig. 5.** Stress distributions ahead of the crack for the SET specimen for different values of the strain hardening exponent $N$: (a) influence of the remote load $K_I/\sigma_y\sqrt{l}$, and (b) influence of the $l/R_0$ ratio.

Figure 5 represents the effect of the remote load $K_I/\sigma_y\sqrt{l}$ on the effective normalized stress $\sigma_e/\sigma_Y$ distribution ahead of the crack tip in the SET specimen. In Fig. 5a, the applied stress intensity factors are $\bar{K}_1 =6.71$ and $\bar{K}_1 =30.58$ with the specified intrinsic parameter value $l =5$ μm, and the plastic work hardening exponent being equal to $N =0.1$, 0.2, and 0.5. The stress level increases with the applied load, in agreement with expectations, and both the plastic zone and the gradient-dominated region augment in size by raising $K_I$. The trends observed in Fig. 4 persist for different $K_I$ levels; the singularity order exceeds that of linear elasticity and the asymptotic crack tip fields appear to be sensitive to the strain hardening exponent. Figure 5b shows the dimensionless effective stress relative to the normalized distance to the crack tip ahead of the crack tip for three normalized values of the internal material length $l/R_0$=0.0087, 0.043, and 0.087 with the work hardening exponent being equal to $N$ =0.075 and 0.5. For the case of $N = 0.5$, the stress distribution follows the $r^{-1/2}$ singularity order of linear elastic solids. A more singular behavior is observed for the case of $N = 0.075$; the numerical solution is not as smooth as for $N = 0.5$ but for the three load levels a higher slope can be observed. Also, the slope appears to be similar, suggesting a solution of similar nature to that of the HRR field, with the singular order being a constant value for a fixed value of $N$.

Further insight is gained by inspecting the angular variation of the stress distributions in the SET specimen. For this purpose, a dimensionless effective stress can be defined as:

$$S_e = \frac{\sigma_e(\theta)}{\sigma_{e,max}} \tag{16}$$

Dimensionless plots of $S_e$ can be shown by considering $x = S_e \cos\theta$ and $y = S_e \cos\theta$. First, Fig. 6 shows the angular stress variations at different $r/l$ values. Angular stress plots are shown for five different values of the strain hardening exponent $N$ ranging from 0.1 to 0.5. The applied stress intensity factor is $\bar{K}_1 = 21.62$, the intrinsic parameter value is $l$ =5 μm, the applied nominal stress level is $\sigma^\infty/\sigma_Y$ =0.155, and the plastic work hardening exponents are $N$ =0.1, 0.2, 0.3, and 0.5. Fig. 6 shows how the contour plots of the dimensionless effective stresses change shape and increase in size with a gradual increase in the crack tip distance $r/l$. Far from the crack tip, close to the elastic-plastic boundary ($r/l = 5$, Fig. 6d), the angular stress distributions show a very small sensitivity to the strain hardening exponent $N$, in agreement with expectations. Noticeable differences are shown as we approach the crack tip, with Fig. 6c ($r/l = 0.5$) showing similar angular distributions in the range $N = 0.1 - 0.4$ but a very different response to the case $N = 0.5$. This behavior agrees with the results observed for $\theta = 0°$ (see, e.g., Fig. 4), as the $N = 0.5$ case shows a stress distribution close to that predicted by linear elasticity while for $N < 0.5$ the stress distributions follow that of conventional plasticity before local strengthening due to strain gradient effects is observed.

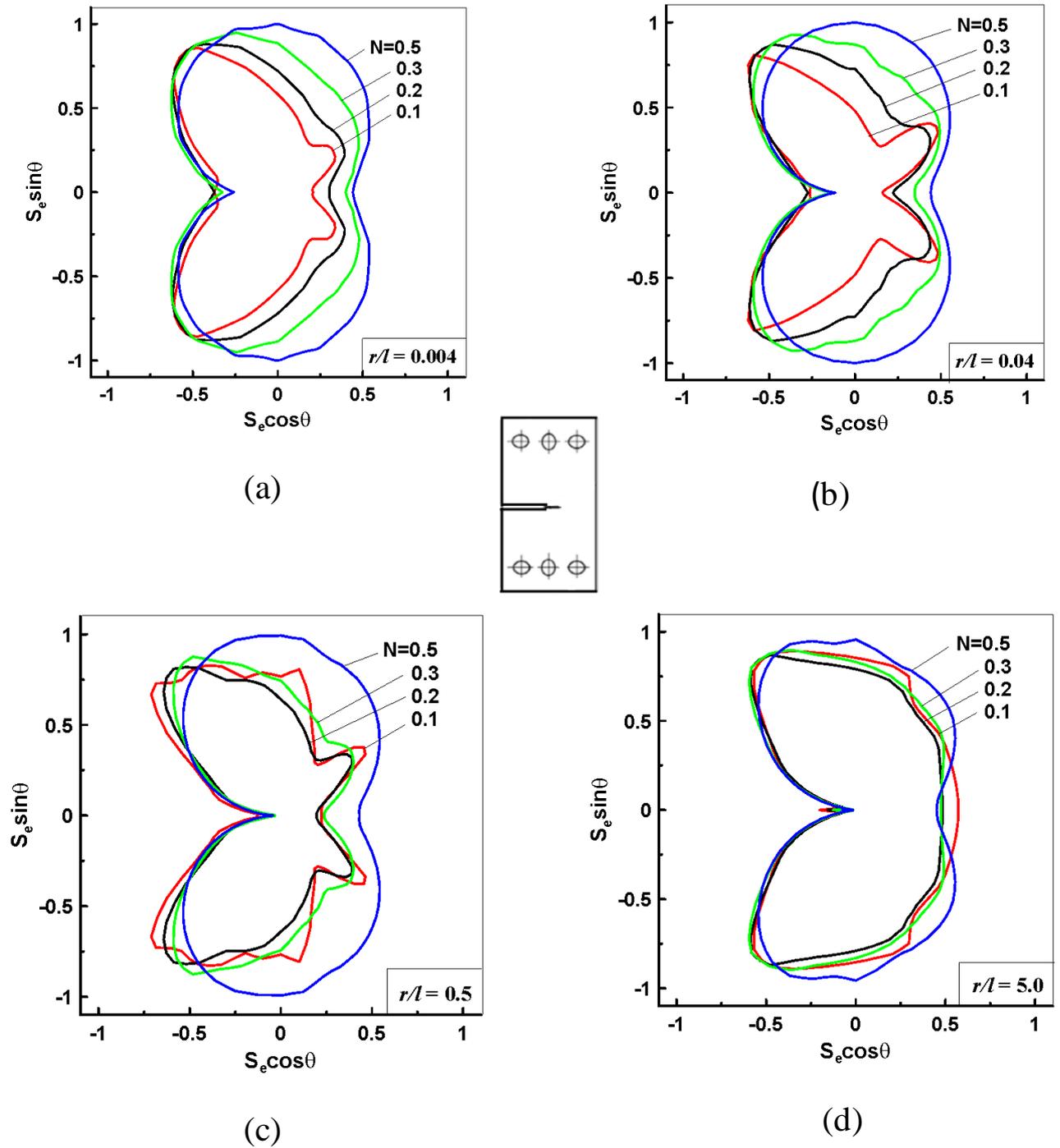

**Fig. 6.** Angular distributions of the normalized effective stress predicted by CMSG plasticity as a function of the strain hardening exponent and different locations ahead of the crack tip: (a) $r/l = 0.004$, (b) $r/l = 0.04$, (c) $r/l = 0.5$, (d) $r/l = 5$. Remote load $\sigma^{\infty}/\sigma_Y = 0.155$. Material properties: $\sigma_Y/E = 0.1\%$, $\nu = 0.3$ and $l = 5\ \mu m$.

Within the strain gradient dominated zone ($r/l < 0.1$, Figs. 6a and 6b) the angular distributions of the normalized effective stress show notable sensitivity to the value of $N$, revealing again a sensitivity of the stress field close to the crack tip to the plastic properties of the material. This is further shown in Fig. 7, where it can be seen that angular stress distributions are almost insensitive to the ratio $r/l$ for $N = 0.5$ but show noticeable differences for $N < 0.5$. It is evident from Figs. 7a-7c that the angular stress distribution in the singularity dominated region is sensitive to the value of the strain hardening exponent. This result would suggest that either the GND term in Eq. (8) is on the same order as the strain related term, or that the strain gradient term $l\eta^p$ is sensitive to the strain hardening exponent.

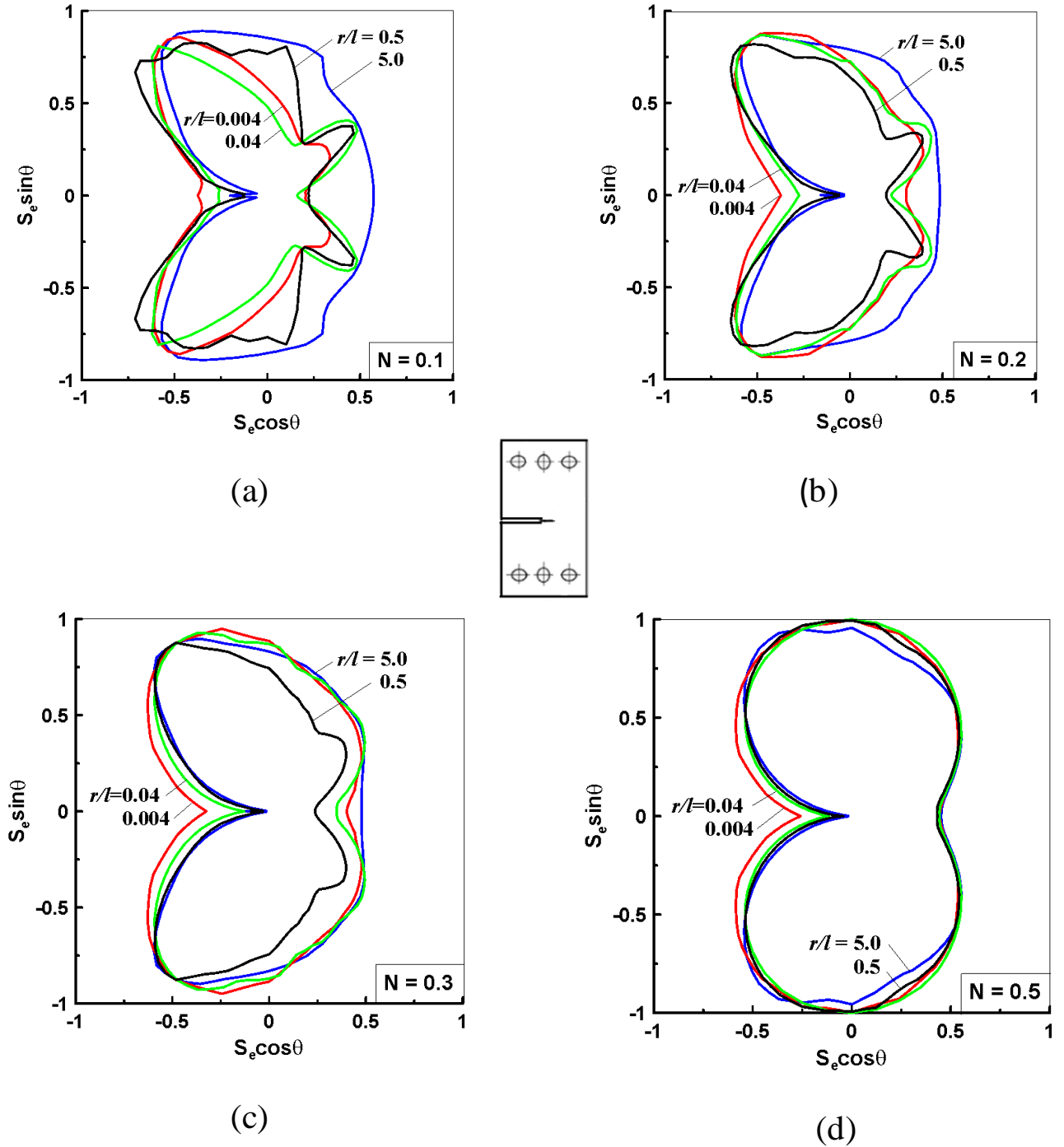

**Fig. 7.** Angular distributions of the normalized effective stress predicted by CMSG plasticity as a function of the distance ahead of the crack tip and strain hardening exponent: (a) $N = 0.1$, (b) $N = 0.2$, (c) $N = 0.3$, $N = 0.5$. Remote load $\sigma^\infty/\sigma_Y = 0.155$. Material properties: $\sigma_Y/E = 0.1\%$, $\nu = 0.3$ and $l = 5\,\mu m$.

We proceed to use the biaxially loaded plate (BLP) and a Ramberg-Osgood hardening law to establish a direct comparison with the analytical HRR solution. Recall that a plate under equibiaxial tension has a $T$-stress equal to zero. Two values of the strain hardening exponent are considered $N = 0.1$ and $N = 0.5$, so as to showcase the influence of material hardening on the angular stress distribution. As with the SET sample results, the normalized effective stress $S_e$ is shown at selected $r/l$ ratios: 0.004, 0.06 and 0.5. The results are shown in Fig. 8.

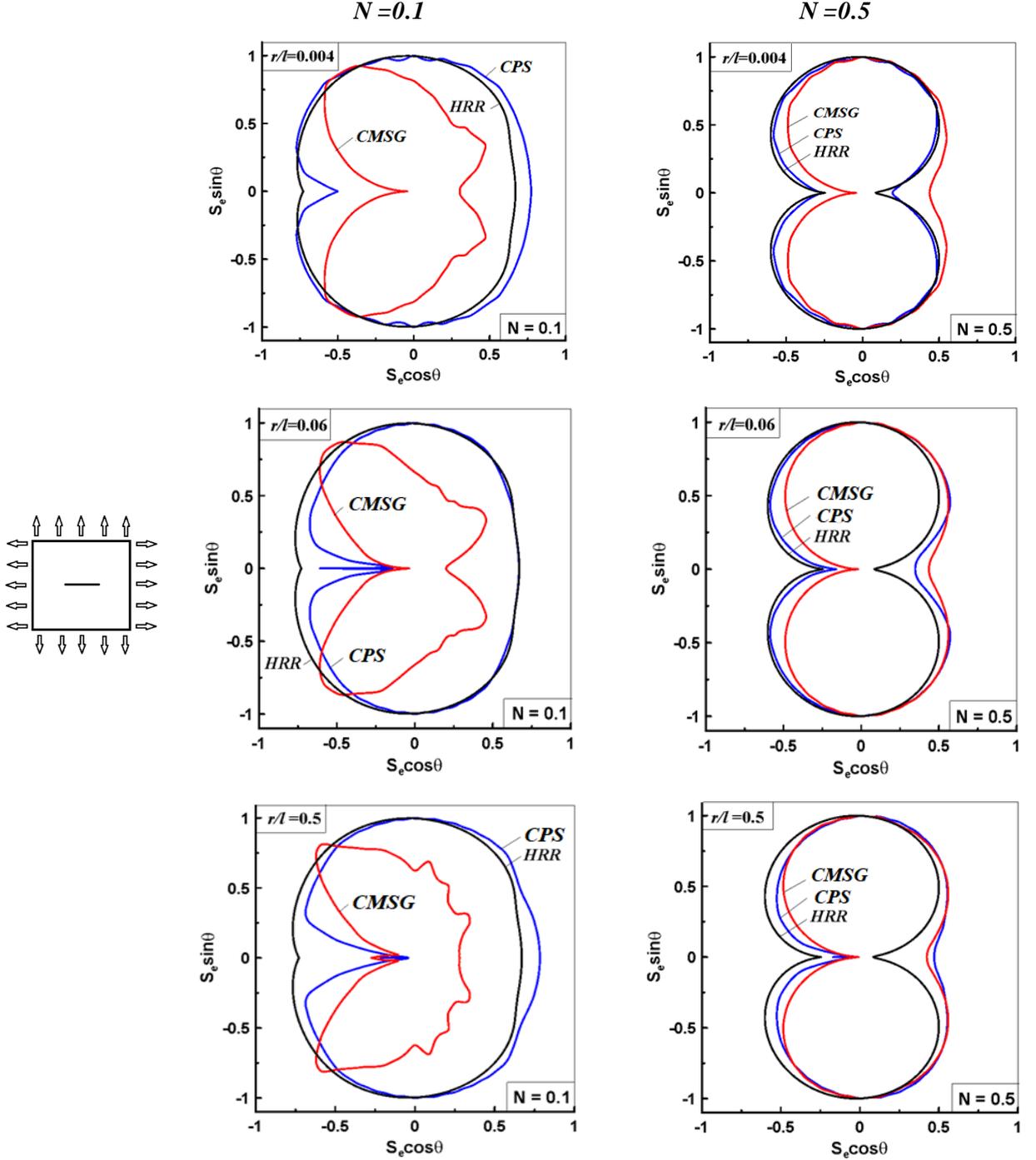

**Fig. 8.** Angular distributions of the normalized effective stress in the BLP specimen for CMSG plasticity, conventional plasticity and the HRR solution. Results are shown for two values of the strain hardening exponent ($N = 0.1, 0.5$) and three $r/l$ ratios (0.004, 0.06 and 0.5). Remote load $\sigma^\infty/\sigma_Y = 0.155$. Material properties: $\sigma_Y/E = 0.1\%$, $\nu = 0.3$ and $l = 5\ \mu m$.

Three constitutive models are considered. The CMSG plasticity computations using the constitutive model described in Section 2 are denoted as CMSG; the finite element results obtained using conventional J2 flow theory with a Ramberg-Osgood hardening law are denoted as CPS; and the analytical HRR (Hutchinson, 1968; Rice and Rosengren, 1968) solution is denoted as HRR.

The results reveal a very good agreement between the J2 flow theory results and the analytical HRR solution, in agreement with expectations. Also, little differences between models are observed when the strain hardening exponent is equal to $N = 0.5$ or higher, as the solution is close to the elastic one. However, notable differences are observed between conventional plasticity and CMSG plasticity for the case of $N = 0.1$. Unlike the conventional plasticity case (both HRR and our finite element results), the CMSG plasticity-based analysis shows significant differences with increasing the $r/l$ ratio.

The sensitivity of the angular stress distributions to the plastic material properties is investigated in Fig. 9. Both the cases of CMSG plasticity and conventional J2 flow theory are considered. The calculated angular distributions of hoop stresses $\hat{\sigma}_\theta^{FEM}(r,\theta,N)$, with the normalization condition $\hat{\sigma}_{e,\max}^{FEM} = \left(\frac{3}{2} S_{ij}^{FEM} S_{ij}^{FEM}\right)_{\max}^{1/2} = 1$, are plotted in Fig. 9 for hardening exponents of $N = 0.1, 0.2, 0.3,$ and $0.5$. The applied stress intensity factor is $\bar{K}_1 = 21.62$, the intrinsic parameter value is $l = 5$ μm, and the applied nominal stress level is $\sigma/\sigma_Y = 0.155$. The difference between dimensionless stress fields in strain-gradient plasticity is significant, especially for N=0.1, and this difference gradually disappears with an increase in the degree of hardening in the order of transition from plasticity to elasticity.

The comparison of these $\hat{\sigma}_\theta^{FEM}(r,\theta,N)$ variations to each other as a function of the radial coordinate $r/l$ leaves little doubt that the strain gradient plasticity fields produced in this section are not of separable form (except for $N \geq 0.5$), i.e.,

$$\bar{\sigma}_{ij}^{CMSG}(r,\theta) \neq \frac{A}{r^\lambda} \tilde{\sigma}_{ij}(\theta), \tag{17}$$

where $(r,\theta)$ are the polar coordinates centered at the crack tip, $\sigma_{ij}$ is the stress and $\tilde{\sigma}_{ij}(\theta)$ the corresponding angular function, $\lambda$ is the power of stress singularity, and $A$ is the amplitude factor depending on the applied loading, cracked body configuration, and material properties. Similar dependences of the angular distributions on the radial coordinate have been found in the analytical and numerical results of Xia and Hutchinson (1996) for Mode I and II crack tip fields for plane strain deformations of an elastic-plastic solid whose constitutive behavior is described by a phenomenological, higher-order strain gradient plasticity theory.

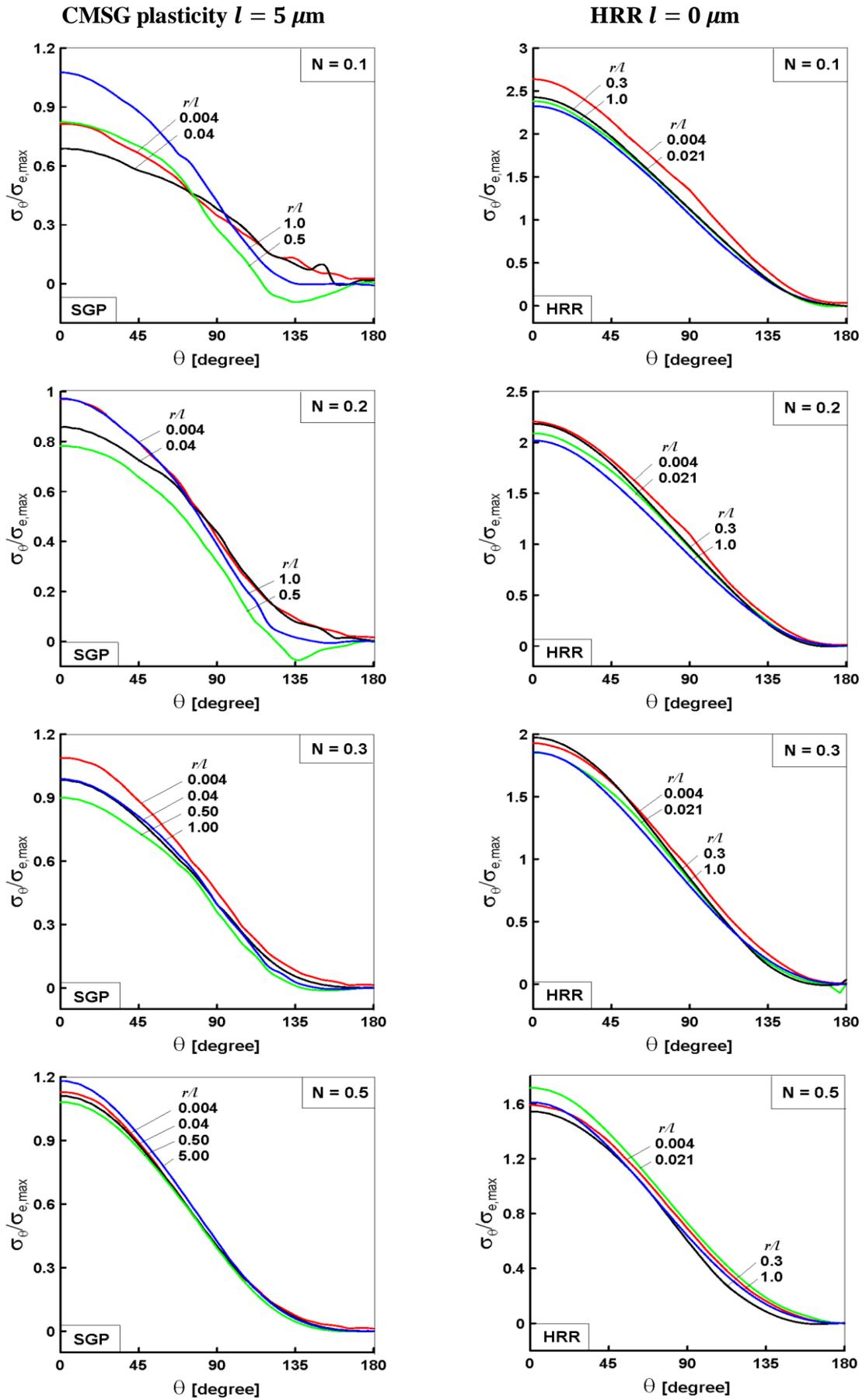

**Fig. 9.** CMSG plasticity and HRR hoop stress angular distributions as a function of $N$ for the SET specimen. Remote load $\sigma^\infty/\sigma_Y = 0.155$. Material properties: $\sigma_Y/E = 0.1\%$, $\nu = 0.3$, $l = 5\ \mu m$.

This is in remarkable contrast to the results obtained for $l = 0$, conventional J2 flow theory. The results denoted as HRR in Fig. 9 indicate that for the same loading conditions, the normalized stress distributions depend only on the polar angle $\theta$ and the strain hardening exponent but are roughly insensitive to the crack tip distance $r/l$ in the range 0.004 to 1.0. In other words, the computational results shown in Fig. 9 confirm the structure of the corresponding HRR field, which is the separable or self-similar asymptotic singular crack tip fields:

$$\bar{\sigma}_{ij}^{HRR}(r,\theta) = K_P r^{\frac{-1}{n+1}} \tilde{\sigma}_{ij}(\theta), \tag{18}$$

where $n=1/N$ is the power of stress singularity, and $K_P$ is the plastic stress intensity factor (Hilton and Hutchinson, 1971).

## 5. Formulation of fracture resistance parameters

We proceed to formulate new fracture resistance parameters for CMSG plasticity. We distinguish between amplitude coefficients and plastic stress intensity factors by making use of both numerical and analytical formulations. In the framework of the analytical solution of the problem, we must connect the structure of the fields at the crack tip with a governing parameter in the form of a suitable path-independent $J$-integral.

*5.1. FEM-amplitude and stress intensity factors.*

Let us consider the structure of the numerical (FEM) crack tip stress fields to be of the following form

$$\bar{\sigma}_{ij}^{FEM}(r,\theta) = A_P^{FEM}(r,\theta=0) \hat{\sigma}_{ij}^{FEM}(r,\theta), \tag{19}$$

where $A_P^{FEM}(r,\theta=0)$ is denoted as quantitative, and $\hat{\sigma}_{ij}^{FEM}(r,\theta)$ is the qualitative component of the general solution (19). The angular distributions of the qualitative component $\hat{\sigma}_{ij}^{FEM}(r,\theta)$ are normalized, such that $\hat{\sigma}_{\theta\theta}^{FEM}|_{\theta=0}=1$ or $\hat{\sigma}_{e,\max}^{FEM} = \left(\frac{3}{2} S_{ij}^{FEM} S_{ij}^{FEM}\right)_{\max}^{1/2} = 1$. Solving equation (19) with respect to the amplitude factor, we obtain

$$A_P^{FEM}(r,\theta=0) = \frac{\bar{\sigma}_{ij}^{FEM}(r,\theta)}{\hat{\sigma}_{ij}^{FEM}(r,\theta)} \tag{20}$$

For the purpose of comparison with the stress field structures at the crack tip known in the literature, we assume that

$$A_P^{FEM} = K_P^{FEM} \bar{r}^\gamma, \gamma < 0 \qquad (21)$$

and

$$K_P^{FEM} = A_P^{FEM} / \bar{r}^\gamma \qquad (22)$$

where $\bar{r} = r/l$ is the normalized crack tip distance, $\gamma$ is the power of stress singularity, and $A_P^{FEM}(r,\theta)$ and $K_P^{FEM}$ are the amplitude and plastic stress intensity factors, respectively, depending on the applied loading, cracked body configuration, and material properties. The structure of the crack tip stress fields in Eq. (19) agrees with that found by Xia and Hutchinson (1996) for the deformation theory of phenomenological higher order strain gradient plasticity model employed in their work, which combines the symmetric part of the Cauchy stress, the asymmetric coupled stress, and the deformation curvature tensor. Their asymptotic analysis generates two amplitude factors, one is the classical HRR solution plastic stress intensity factor $K_P = (J/(\sigma_0 \varepsilon_0 I_n))^{n/(n+1)}$, and the second, $A$, is related to the dimensionless $\theta$-variations for dominantly stress components $\hat{\sigma}_{ij}, \hat{\tau}_{ij}$. Apart from the amplitude factors, Xia and Hutchinson (1996) introduced the dominantly singular crack tip strain and stress fields, where the dimensionless angular functions $\hat{\varepsilon}_{ij}, \hat{\chi}_{ij}, \hat{\sigma}_{ij}, \hat{\tau}_{ij}, \hat{m}_{ij}$ depend on the polar angle $\theta$, strain hardening exponent $n$, and the normalized $\bar{r} = r/l$ length parameter, characterizing the scale across which gradient effects become important. The authors show that in the outer field where the HRR solution is approached, $\hat{\varepsilon}_{ij}$ and $\hat{\sigma}_{ij}$ approach $\tilde{\varepsilon}_{ij}$ and $\tilde{\sigma}_{ij}$, respectively, defined by Eq.(18) in the HRR theory.

First, we proceed to identify fracture parameters based on the numerical results, as computed for the loading conditions and material properties listed in Section 3. Recall that Fig. 4 shows the normalized stress distribution ahead of the crack tip ($\theta = 0°$) for both CMSG plasticity and conventional plasticity. Focusing on the CMSG plasticity results, it is evident that the singular behavior of the asymptotic crack tip fields is the result of the combined influence of plastic strain gradients and plastic material properties of the material. In other words, unlike previous works, we see a sensitivity to the material strain hardening exponent $N$; this is quantified by fitting the numerical results in Fig. 4 such that the power of the stress singularity is given by the following equations:

$$\text{for SET: } \gamma = 1.08333 N^3 - 0.76785714 N^2 + 0.3323809524 N - 0.6196 \qquad (23)$$

$$\text{for BLP: } \gamma = 0.0333 N^3 - 0.2193 N^2 + 0.0517 N - 0.6816 \qquad (24)$$

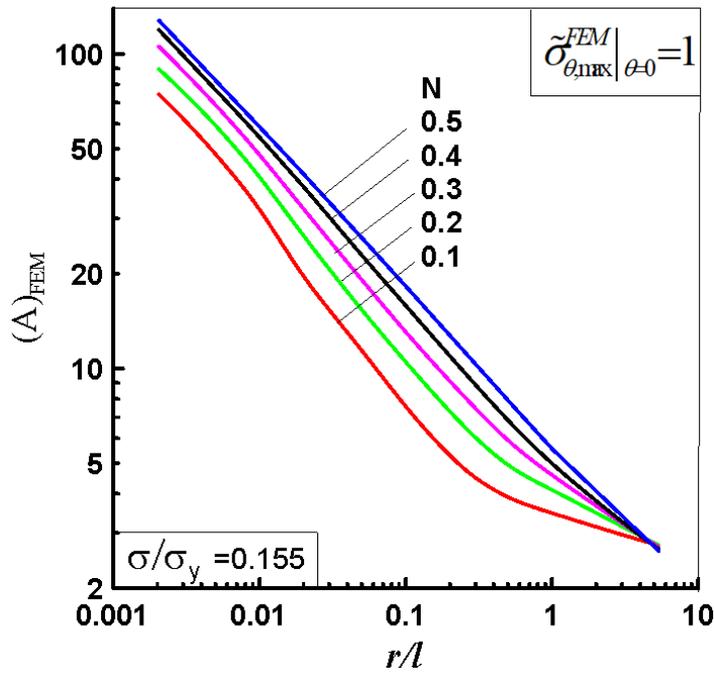
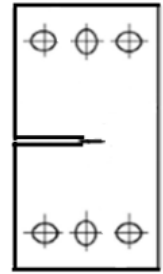
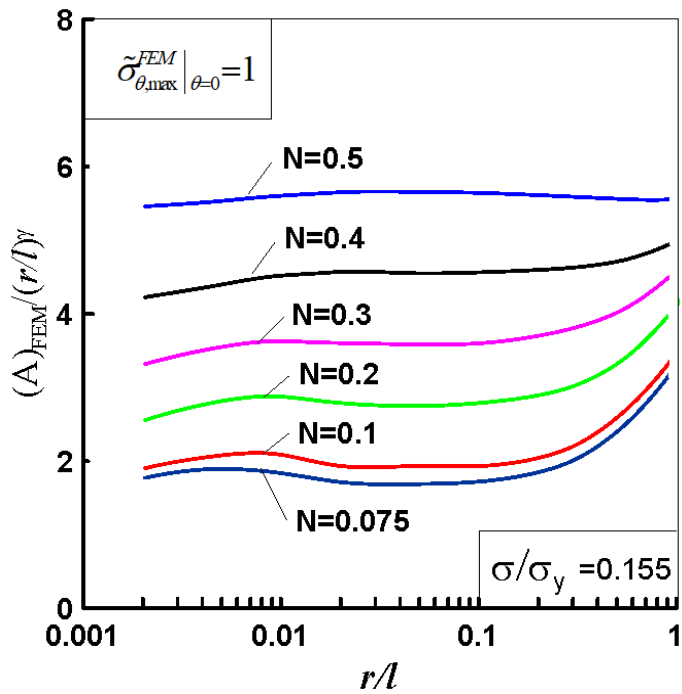

**Fig. 10.** CMSG plasticity FEM-amplitude and stress intensity factors behavior in the SET specimen. Remote load $\sigma^\infty/\sigma_Y = 0.155$. Material properties: $\sigma_Y/E = 0.1\%$, $\nu = 0.3$ and $l = 5\ \mu m$.

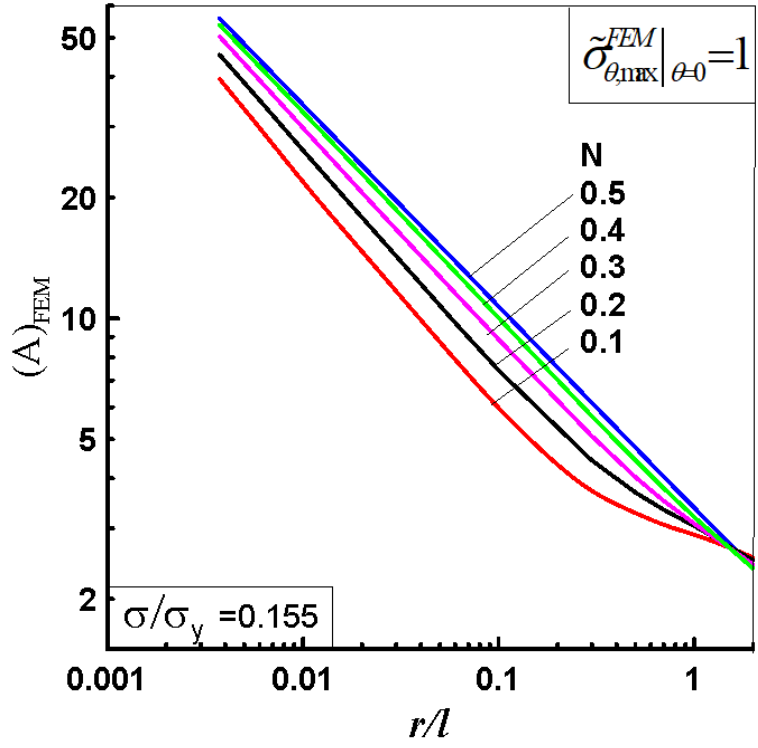

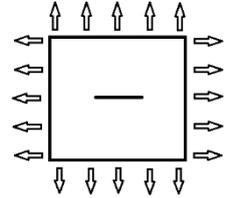

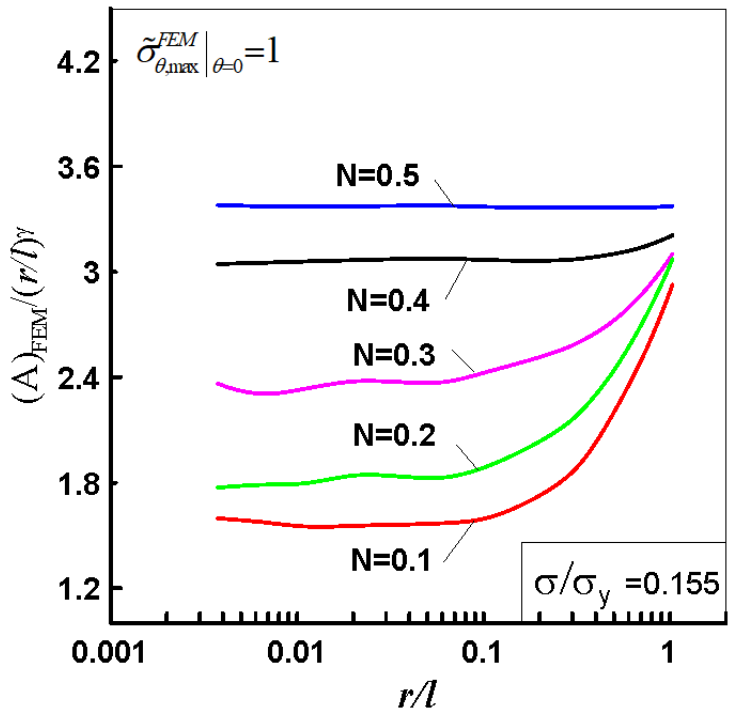

**Fig. 11.** CMSG plasticity FEM-amplitude and stress intensity factors behavior in the BLP specimen. Remote load $\sigma^\infty/\sigma_Y = 0.155$. Material properties: $\sigma_Y/E = 0.1\%$, $\nu = 0.3$ and $l = 5\,\mu$m.

The numerical amplitude factor $A^{FEM}$ and the plastic stress intensity factor $K_P^{FEM}$ are shown in Fig. 10 for the case of the SET sample as a function of the distance ahead of the crack tip $r/l$ based on Eqs. (20), (23)-(24). The results reveal that the numerical values for amplitude factor $A_P^{FEM}$ significantly decrease as a function of the normalized crack tip distance $r/l$, and all separate curves for different $N$ coincide with each other when $r/l>2$. In contrast, when the nondimensional distance to the crack tip is less than $0.1 r/l$, the behavior of the plastic stress intensity factor $K_P^{FEM} = A_P^{FEM} / \bar{r}^\gamma$ is almost uniform with respect to these dimensionless quantities for $r/l$ ranging from 0.001 to 0.1 for each value of the work hardening exponent $N$. This region of the steady state strain gradient effects may be regarded as a domain of determination or domain of validity for the CMSG plasticity. Our numerical analysis confirms the results of Xia and Hutchinson's (1996) finite element studies, which clearly show that HRR field is accurate for $r/l >5$, the singular field for strain-gradient plasticity is dominant for $r/l <5$, and a gradual transition region lies in between. The dominant zone size remains significantly larger than dislocation spacing, such that the use of continuum plasticity is justified. The same trends are observed for the BLP specimen, as shown in Fig. 11.

*5.2. Amplitude and stress intensity factors based on asymptotic crack tip fields.*

This study uses the conventional theory of the mechanism-based strain gradient (CMSG) plasticity detailed by Huang et al. (2004), which is a lower-order theory based on the Taylor dislocation model. The solid is assumed to be homogeneous and isotropic with an energy density $W$, which depends on scalar invariants of the strain tensor, while the deformation curvature tensor and coupled stress are negligible. Our starting point is to look for a dominant singular solution derived from an incompressible, irrotational displacement field.

Numerous previous asymptotic fields around crack tips have separable forms of solutions, such as the classical $K$ field, HRR field (1968), crack tip field in the couple-stress theory of strain gradient plasticity (Huang et al., 1995), as well as the crack tip field in phenomenological theory of strain gradient plasticity (Fleck and Hutchinson,1997). These fields are all governed by the path-independent $J$-integral. To formulate the problem for the asymptotic strain-gradient crack tip fields, Chen et. al. (1999) and Shi et. al. (2000, 2001) employed a displacement potential $\phi$ similar to the HRR field in classical plasticity. A separable form for $\phi$ is taken as

$$\phi = r^p \tilde{\phi}(\theta), \tag{25}$$

where $p$ is the power of stress singularity ($p>0$) and $\tilde{\phi}(\theta)$ is the angular distribution of $\phi$. Huang et al. (1995) and Xia and Hutchinson (1996) analytically obtained the asymptotic fields near a Mode I crack tip in rotation-gradient-based strain gradient plasticity. They established that the crack tip deformation field is irrotational, such that the strains and stresses are more singular than curvatures (strain gradients) and higher-order stresses, respectively. The corresponding power $p$ for displacements in this stress-dominated crack tip field, where stresses are more singular than higher-order stresses, is $p=1/(n+1)$, which is the same as for the HRR field in classical plasticity.

Shi et. al. (2001) observed that the stress field in MSG plasticity is more singular than both the HRR field and the classical elastic $K$ field ($r^{-0.5}$), which is consistent with Jiang et al.'s (2001) finite element analysis. They found that the Mode I stress singularity in MSG plasticity, $\lambda_I= 0.63837$, is slightly smaller than that for Mode III, $\lambda_{III}= 0.65717$. Even though this was not observed in classical elastic or elastic-plastic crack tip fields, similar observations have been made in other strain gradient plasticity theories. For example, Huang et al. (1995, 1997) and Xia and Hutchinson (1996) showed that the stress traction ahead of a Mode I crack tip predicted by Fleck and Hutchinson's couple-stress theory of strain gradient plasticity (Fleck and Hutchinson, 1993; Fleck et al., 1994) has the same singularity as the field. Shi et. al. (2001) pointed out that, unlike the HRR field in classical plasticity, the power of stress singularity in MSG plasticity is independent of the plastic work hardening exponent $N$. The authors argued that this is because the strain gradient becomes more singular than the strain near the crack tip and dominates the contribution to the flow stress. This indicates that the density $\rho_G$ of GND around a crack tip is significantly larger than the density $\rho_S$ of SSDs.

Unlike these separable asymptotic crack tip fields previously established, Xia and Hutchinson (1996) used the dominantly singular crack tip strain and stress fields within the framework of coupled stress theory, where the dimensionless angular functions $\hat{\varepsilon}_{ij}, \hat{\chi}_{ij}, \hat{\sigma}_{ij}, \hat{\tau}_{ij}, \hat{m}_{ij}$ depend on the polar angle $\theta$, strain hardening exponent $n$ and the normalized $\bar{r} = r/l$ the intrinsic material length parameter, i.e.,

$$\left[\sigma_{ij}, l^{-1}m_{ij}, \tau_{ij}\right] = \sigma_0 \left(J/(\sigma_0\varepsilon_0 I_n r)\right)^{n/(n+1)} \left[\hat{\sigma}_{ij}\left(\theta, \frac{r}{l}, n\right), \hat{m}_{ij}\left(\theta, \frac{r}{l}, n\right), \hat{\tau}_{ij}\left(\theta, \frac{r}{l}, n\right)\right] \quad (26)$$

In the present study, following Xia and Hutchinson (1996) and Shi et. al. (2001), the asymptotic displacement and stress dominated fields near a crack tip are written in the context of CMSG plasticity as:

$$\bar{u}_i(r,\theta) = \alpha K_P^{1/N} \bar{r}^{(2-2\lambda)} \hat{u}_i(r,\theta), \quad i=1,2 \quad (27)$$

$$\bar{\sigma}_{\alpha\beta}(r,\theta) = K_P \bar{r}^{-\lambda} \hat{\sigma}_{\alpha\beta}(r,\theta) \quad (28)$$

$$\bar{\sigma}_e(r,\theta) = K_P \bar{r}^{-\lambda} \hat{\sigma}_e(r,\theta), \quad (29)$$

where $\bar{r} = r/L$, and $L$ is a characteristic length (e.g., the strain gradient length scale). In Eqs.(24)–(26), $K_P$ denotes the plastic stress intensity factor, $N$ is the strain hardening exponent, and $\lambda > 0$ is the power of stress singularity.

We proceed then to define a suitable J-integral (Rice, 1968) for CMSG plasticity. Suitable path-independent J-integrals for materials with strain-gradient effects have been developed by Huang et al. (1995, 1997), Xia and Hutchinson (1995, 1996), Chen et al. (1999), and Martinez-Paneda and Fleck (2019). Most general forms for the high-order theories can be written as

$$J = \int_\Gamma \left( W n_1 - T_i u_{i,1} - q_i \omega_{i,1} \right) dS, \qquad (30)$$

where $W$ is the strain energy density, $u_i$, $\omega_i$, $T_i$, and $q_i$ are displacements, rotations, tractions, and couple-stress tractions, respectively, and $n_1$ is the unit normal. As mentioned above, Xia and Hutchinson (1996) employed two amplitude factors to formulate the problem for the crack tip fields; one is the classical HRR solution plastic stress intensity factor $K_P$, and the second, $A$, is related to the dimensionless $\theta$-variations for dominant stress components $\hat{\sigma}_{ij}, \hat{\tau}_{ij}$. Huang et al. (1997) introduced several definitions for amplitude factors for the asymptotic crack tip fields. Namely, in an elastic material with strain-gradient effects, the dominant stress and couple-stress fields in Mode I are governed by two independent parameters, $B_I^{(0)}$ for stresses (similar to $K_I$ in classical elastic fracture mechanics) and $A_I^{(0)}$ for couple stresses. In the cases of an elastic-plastic strain-gradient material with the separated or integrated law of hardening, the authors used $B_I^{(0)}$ and $A_I^{(0)}$ amplitude factors to conjugate with additional nondimensional functions $I_1(n)$ and $I_2(n)$ of the hardening power $n$.

In the case of CMSG plasticity, because it does not involve higher-order stresses, one can use as a starting point the J-integral definition provided by Rice (1968),

$$\bar{J} = \int_\Gamma \left( \bar{W} dy - \bar{\sigma}_{ij} n_j \bar{u}_{i,x} d\bar{s} \right). \qquad (31)$$

while noting that stresses, strains and displacements are given by the constitutive behavior of CMSG plasticity.

For convenience, we introduce the following non-dimensional quantities:

$$\bar{J} = \frac{JE}{\sigma_y^2}, \quad \bar{W} = \frac{WE}{\sigma_y^2}, \quad \bar{\sigma}_{ij} = \frac{\sigma_{ij}}{\sigma_y}, \quad \bar{u}_i = \frac{u_i E}{\sigma_y}, \qquad (32)$$

where $\sigma_y$ is the yield stress.

The strain energy density in Eq. (31) is defined as the sum of elastic and plastic components:

$$\bar{W} = \bar{W}_E + \bar{W}_P = \left[\frac{1+\nu}{3}\bar{\sigma}_e^2 + \frac{1-2\nu}{6}\bar{\sigma}_{kk}^2\right] + \left[\frac{\alpha}{N+1}\bar{\sigma}_e^{\frac{N+1}{N}}\right]. \tag{33}$$

The value of $J$-integral for elastic Mode I far from crack tip stress fields is defined as

$$J_C = \frac{K_1^2}{E}, \quad \bar{J}_C = \frac{J_C E}{\sigma_y^2 L} = \frac{1}{L}\left(\frac{K_1}{\sigma_y}\right)^2 \tag{34}$$

Conveniently, the asymptotic crack tip field can be used to evaluate the $J$-integral. Substituting Eqs.(27)–(29) and Eq.(33) into Eq.(31) we obtain the following line $J$-integral

$$\bar{J} = \int_{-\pi}^{+\pi} \left(\bar{W} r \cos\theta d\theta - \bar{\sigma}_{ij} n_j \bar{u}_{i,x} r d\theta\right) =$$

$$= r\left\{\left(\frac{K_1}{\sigma_y \sqrt{r}}\right)^2 \int_{-\pi}^{+\pi}\left[\frac{1+\nu}{3}(\tilde{\sigma}_e^{el})^2 + \frac{1-2\nu}{6}(\tilde{\sigma}_{kk}^{el})^2\right]\cos\theta d\theta + \frac{\alpha K_P^{\frac{1+N}{N}}}{N+1}\int_{-\pi}^{+\pi}(\bar{r}^{-\lambda}\hat{\sigma}_e)^{\frac{N+1}{N}}\cos\theta d\theta\right\} -$$

$$-\alpha K_P^{\frac{1+N}{N}}\int_{-\pi}^{+\pi}\left\{\begin{array}{l}\left[\bar{r}^{(2-3\lambda)}(\hat{\sigma}_{rr}\hat{u}_\theta + \hat{\sigma}_{\theta r}\hat{u}_r)\cos\theta - \bar{r}^{(1-3\lambda)}\left(\hat{\sigma}_{rr}\frac{\partial \hat{u}_r}{\partial \theta} + \hat{\sigma}_{\theta r}\frac{\partial \hat{u}_\theta}{\partial \theta}\right)\sin\theta\right] + \\ (2-2\lambda)\bar{r}^{(1-3\lambda)}(\hat{\sigma}_{rr}\hat{u}_r + \hat{\sigma}_{r\theta}\hat{u}_\theta)\cos\theta\end{array}\right\} r d\theta \tag{35}$$

The path-independence of the $J$-integral allows the equality between Eq. (34) and Eq. (35) to be written as

$$\bar{J}_C = \bar{J}. \tag{36}$$

Equation (36) can be rewritten in the following form by considering Equations (34)-(35)

$$\left(\frac{\bar{K}_1}{\sigma_y}\right)^2\left(\frac{a}{r}\right)\left[1-\frac{I_{el}}{2\pi}\right] = \alpha(K_P)^{\frac{1+N}{N}}\left[\frac{1}{N+1}\bar{r}^{\frac{-\lambda(1+N)}{N}}I_{n,1} - \bar{r}^{(2-3\lambda)}I_{n,2} + \bar{r}^{(1-3\lambda)}I_{n,3} - (2-2\lambda)\bar{r}^{(1-3\lambda)}I_{n,4}\right], \tag{37}$$

where

$$\bar{K}_1 = \left(K_1/\sqrt{a}\right) \tag{38}$$

$$I_{el} = \int_{-\pi}^{+\pi} f(\tilde{\sigma}_{ij}^{el}) d\theta \tag{39}$$

$$I_{n,1} = \int_{-\pi}^{+\pi} \hat{\sigma}_e^{\frac{N+1}{N}} \cos\theta d\theta \tag{40}$$

$$I_{n,2} = \int_{-\pi}^{+\pi} (\hat{\sigma}_{rr}\hat{u}_\theta + \hat{\sigma}_{r\theta}\hat{u}_r)\cos\theta d\theta \tag{41}$$

$$I_{n,3} = \int_{-\pi}^{+\pi} \left(\tilde{\sigma}_{rr}\frac{\partial \hat{u}_r}{\partial \theta} + \tilde{\sigma}_{r\theta}\frac{\partial \hat{u}_\theta}{\partial \theta}\right)\sin\theta d\theta \tag{42}$$

$$I_{n,4} = \int_{-\pi}^{+\pi} (\hat{\sigma}_{rr}\hat{u}_r + \hat{\sigma}_{r\theta}\hat{u}_\theta)\cos\theta d\theta \tag{43}$$

The path-independent integral (35) is related to plastic stress intensity factor $K_P$ and four numerical integrals $I_{n,i}(i=1,4)$ represented by Eqs.(40–43). This provides a bridge between this new plastic SIF $K_P$ and the classical elastic SIF. The *J*-integral can be evaluated by Eq. (35) for an elastic-plastic material with strain gradient effects subjected to remotely imposed classical *K*-fields. Details of the derivation of the equation for the elastic part of the strain energy density $W_E$ and the integral $I_{el}$ associated with it are given in Appendix 1. Rearranging Eq. (37) leads to the expression for new plastic stress intensity factor accounting for both the elastic and plastic parts of strain energy density:

$$K_p = \left\{ \frac{\left(\frac{\bar{K}_1}{\sigma_y}\right)^2 \left(\frac{a}{r}\right)\left[1 - \frac{I_{el}}{2\pi}\right]}{\alpha\left[\frac{1}{N+1}\bar{r}^{\frac{-\lambda(1+N)}{N}} I_{n,1} - \bar{r}^{(2-3\lambda)} I_{n,2} + \bar{r}^{(1-3\lambda)} I_{n,3} - (2-2\lambda)\bar{r}^{(1-3\lambda)} I_{n,4}\right]} \right\}^{\frac{N}{N+1}}, \tag{44}$$

where $\bar{r} = r/l$.

A comparison of the terms in Eq. (44) related to numerical integrals $I_{n,i}(i=1,4)$ at the crack tip distances ranging of $(0.001-1.0)\bar{r}$ for all values of the work hardening $N$ shows that the first term $I_{n,1}$ dominates by an order of magnitude or more with respect to all the backward terms $I_{n,i}(i=2,3,4)$, i.e., the contribution of these terms is negligible. Therefore, without losing generality, we may impose the following simplified expression for the plastic stress intensity factor in the asymptotic crack tip problem

$$K_p = \bar{r}^\lambda \left\{ \left(\frac{\bar{K}_1}{\sigma_y}\right)^2 \left(\frac{a}{r}\right)\left(1 - \frac{I_{el}}{2\pi}\right) \bigg/ \left(\frac{\alpha I_{n,1}}{N+1}\right) \right\}^{\frac{N}{N+1}}, \tag{45}$$

and the corresponding expression for the amplitude factor has the form

$$A_P^{asm}(r,\theta) = K_P(\bar{r})^{-\lambda} = \left\{ \left(\frac{\bar{K}_1}{\sigma_y}\right)^2 \left(\frac{a}{r}\right)\left(1 - \frac{I_{el}}{2\pi}\right) \bigg/ \left(\frac{\alpha I_{n,1}}{N+1}\right) \right\}^{\frac{N}{N+1}}. \tag{46}$$

Figure 12 shows the distributions of plastic stress intensity factors $K_P$ according to Eq. (45) for the asymptotic CMSG plasticity problem as a function of the normalized distance in the SET specimen and BL plate. The behavior of the plastic SIFs $K_P$ has approximately the same range of variation as the numerical SIFs $K_P^{FEM}$ (Figs. 10b, 11b), but it exhibits sensitivity to the crack tip distance.

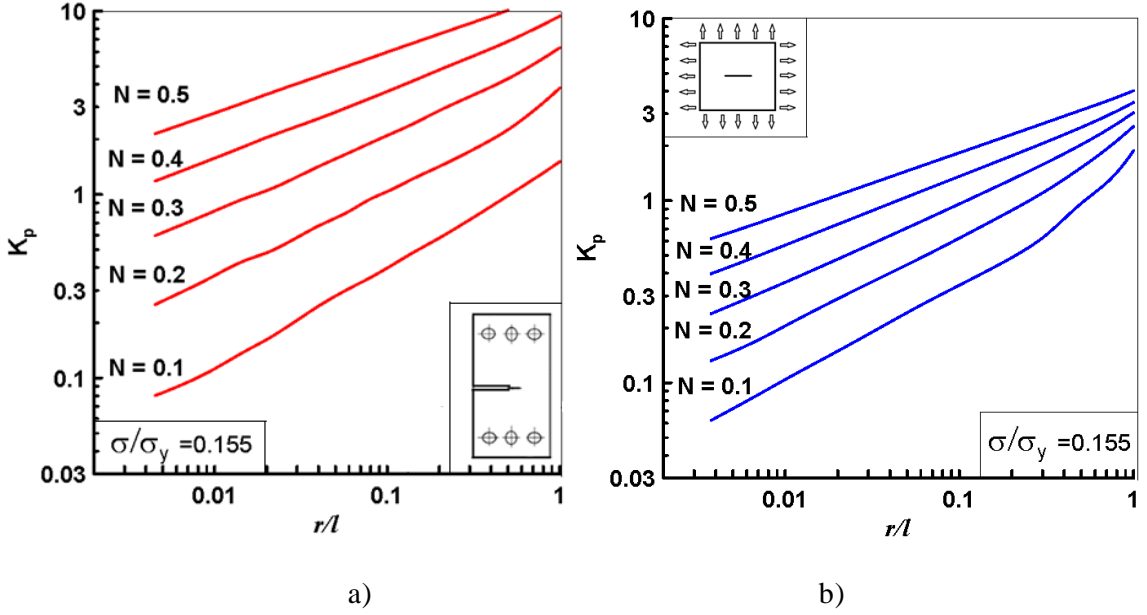

**Fig. 12.** CMSGP asymptotic stress intensity factors behavior in (a) SET specimen and (b) BL plate.

Following Xia and Hutchinson (1996) and Huang et al. (1997), the general amplitude factor $A_P^G(r,\theta)$ is introduced, which combines the asymptotic solution $A_P^{asm}(r,\theta)$ and the second term $A_P^S(r,\theta)$ from the condition for matching the numerical solution. In essence, this approach corresponds to the classical fracture mechanics, when the elastic SIF is the product of the dimensional part $\sigma\sqrt{\pi a}$ and the geometry-dependent factor $Y(a/w)$, i.e., $K_1 = \sigma\sqrt{\pi a} \cdot Y(a/w)$. Thus, the following structure for amplitude factors is introduced to scale with strain gradients from the dimensional consideration:

$$A_P^G(r,\theta=0) = A_P^{asm}(r,\theta=0) \cdot A_P^S(r,\theta=0). \tag{47}$$

Dimension matching requires the following structure for the new amplitude coefficient

$$A_P^G(r,\theta=0) = K_P^G \bar{r}^\delta \tag{48}$$

$$A_P^S(r,\theta=0) = K_P^S \bar{r}^\beta \tag{49}$$

where $K_P^G$ and $K_P^S$ are general and second plastic stress intensity factors, respectively, and $\delta$ and $\beta$ are the corresponding crack tip distance exponents. Figure 13 shows the behavior of $K_P^S$ and $\beta$, which are found from the condition for matching the numerical solution.

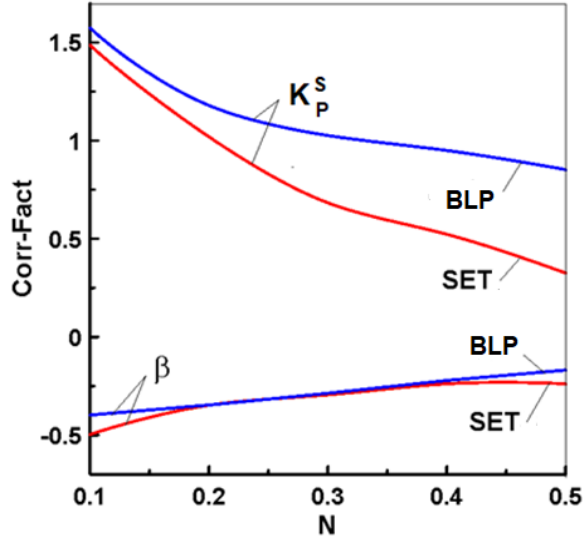

**Fig. 13.** Second SIF $K_P^S$ and $\beta$ power behavior as a function of plastic work hardening exponent N.

By fitting the distributions shown in Fig. 13, the approximation equations that describe the behavior of the second plastic SIF $K_P^S$ and $\beta$ power were determined as a function of plastic work hardening exponent N for the considered cracked bodies

for SET: $K_P^S = 5.1014N^2 - 5.8779N + 2.0093$ (50)

$\beta = -2.1071N^2 + 1.8893N - 0.6586$ (51)

for BLP: $K_P^S = -21.90000 N^3 + 24.50143 N^2 - 9.71586 N + 2.32132$ (52)

$\beta = -1.75000 N^3 + 1.63929 N^2 + 0.13143 N - 0.42560$ (53)

Substituting Eqs. (48,49) into Eq. (47) yield

$$K_P^G \bar{r}^\delta = K_P^S \cdot \bar{r}^\beta \left\{ \left(\frac{\bar{K}_1}{\sigma_y}\right)^2 \left(\frac{a}{r}\right)\left(1 - \frac{I_{el}}{2\pi}\right) \Big/ \left(\frac{\alpha I_{n,1}}{N+1}\right) \right\}^{\frac{N}{N+1}}$$ (54)

and the expression for the general plastic SIF, which will satisfy the condition

$$K_P^G = \left(\bar{r}\right)^{(\beta-\delta)} K_P^S \left\{ \left(\frac{\bar{K}_1}{\sigma_y}\right)^2 \left(\frac{a}{r}\right)\left(1 - \frac{I_{el}}{2\pi}\right) \bigg/ \left(\frac{\alpha I_{n,1}}{N+1}\right) \right\}^{\frac{N}{N+1}}. \qquad (55)$$

The numerical and analytical formulations for the plastic stress intensity factors $K_P^{FEM}$ and $K_P^G$ in the form of equations (22) and (55), respectively, are parameters of fracture resistance, and they are applicable in a domain of determinacy or domain of validity of CMSG plasticity. The proposed plastic stress intensity factor can be considered a reference or generalized parameter (GP) to determine the joint probability of failure assuming a three-parameter Weibull distribution. Such methodology allows the normalized equivalent parameter (GP$_{eq}$) to be proposed for probabilistic/statistical assessment of cleavage fracture in the presence of plastic deformation with a suitable failure variable irrespective of the constraint conditions.

## 6. Discussion

Shi et. al. (2001) found that, unlike the HRR field in classical plasticity, the power of stress singularity in the higher order MSG plasticity theory is independent of the plastic work hardening exponent $N$. However, our results show that for CMSG plasticity, the stress singularity is sensitive to the plastic properties of the material. As shown in Fig. 4, the crack tip stress distribution reveals a similar trend to that observed in the finite element analysis of crack tip fields in MSG plasticity by Jiang et al. (2001); gradient effects elevate the stresses and lead to an asymptotic behavior that appears to be more singular than the linear elastic solution. Also, for both MSG and CMSG plasticity models, the normalized effective stress within the gradient plasticity dominance region increases with increasing plastic work hardening exponent $N$. However, the slope of the stress distribution close to the crack tip appears to be sensitive to changes in $N$, as quantified by Eqs. (23)-(24). As in conventional plasticity, we have shown that a plastic stress intensity factor $K_p$ can be defined, which exhibits a constant value within the region of gradient dominance for a fixed value of $N$ – see Figs. 10b and 11b.

Three annular regions are observed close to the crack tip. An outer elastic region, governed by the remote $K$ field, an elastic-plastic region following the HRR solution and an inner CMSG plasticity region where the stress singularity is equal or higher than the elastic one but dependent on the plastic properties of the material. Shi et al. (2001) investigated the asymptotic crack tip field in the region adjacent to the crack tip and established that the stress dominated asymptotic crack tip field in MSG plasticity has a separable field. The resulting stress singularity is almost constant, around 0.65 for Mode I and Mode III fractures. However, our results (see Fig. 9) show that the crack tip fields of a solid characterized by CMSG plasticity appear to be of non-separable form.

Figure 14 summarizes our findings in the context of the literature, representing a comparison for the crack tip singularity behavior as a function of the plastic work hardening exponent $N$. In this figure, the elastic singularity is given by a straight line 1 with y-axis equal to -0.5. Line 2 with the crack tip singularity $-N/(N+1)$ corresponds to the HRR field in classical plasticity. Line 3 at the bottom shows the behavior of the crack tip singularity with the constant value –0.65 obtained by Shi et al. (2001) as the asymptotic MSG plasticity separable solution. Further, line 4 with the variable crack tip singularity is the numerical result of the present study according the CMSG plasticity. This CMSG prediction lies between the HRR and MSG plasticity lines and exceeds the elastic -1/2 singularity. The behavior of the degrees of singularity for the classical HRR plasticity and the CMSG theory as a function of the work hardening $N$ have opposite trends. This circumstance explains the inverse nature of the arrangement of the curves for plastic stress intensity factors $K_P^{FEM}$ and $K_P$ and their dependence on $N$. It is unclear why the stress singularity of MSG is not sensitive to the strain hardening exponent, unlike its lower-order counterpart. The source of differences could be related to the role that the higher order equilibrium equation plays in suppressing plastic deformation in the vicinity of the crack tip, as observed in other higher order strain gradient plasticity theories (Martínez-Pañeda et al., 2019).

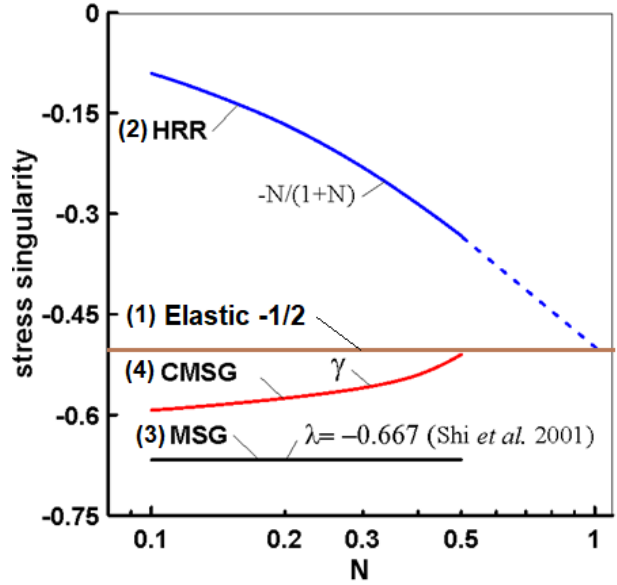

**Fig. 14.** Crack tip stress singularities behavior for HRR, CMSG, and MSG plastic theories. Results obtained with the SET specimen for an external applied load $K_1/\sigma_y \sqrt{l} = 30.58$ ($l$ =5 μm).

The constitutive laws for the CMSG model are one of the simplest generalizations of the $J_2$ deformation theory of plasticity to include strain gradient effects. For an elastic-plastic material, our calculations show that stresses have variable singularity near the crack tip and are governed by the plastic stress intensity factor $K_P^{FEM}$ according to Eqs. (19)–(21). Fig. 15 shows the radial distributions for plastic

SIF $K_P^{FEM}$ for different values of the intrinsic material length $l$ as a function of plastic work hardening exponent $N$. Plastic SIFs are almost independent of the normalized crack tip distance $r/l$ and augment with increasing the material length parameter from 5 to 10 $\mu$m. An important conclusion regarding these numerical results is the sensitivity of the proposed parameter of fracture resistance $K_P^{FEM}$ in the assessment of the coupled effects of the parameters $l$ and $N$, and in the behavior of strain gradient plasticity solids. This property differs from the known elastic and plastic stress intensity factors and is attractive from the point of view of practical applications. The introduced numerical and asymptotical SIFs for gradient plasticity continue a series of new nonlinear stress intensity factors proposed by Shlyannikov et. al. (2014,2015,2018) for the conditions of plasticity, creep, and the creep and fatigue interaction based on the approaches of continuum damage mechanics.

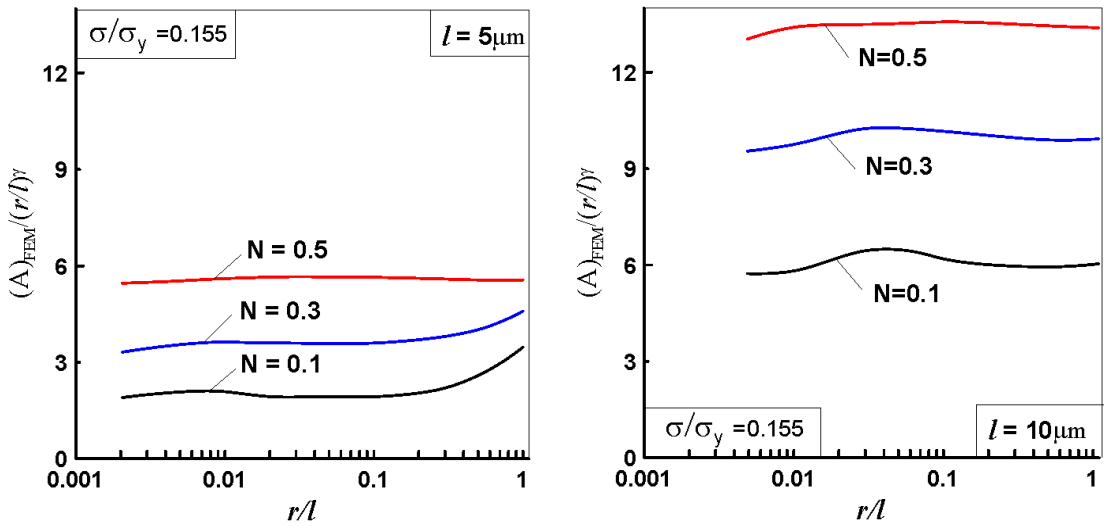

**Fig. 15.** Radial CSMG $K_P^{FEM}$ plastic SIFs distributions for different values of the material length parameter $l$ and the strain hardening exponents $N$.

Experiments such as nano-indentation demonstrate the influence of strain gradients in governing material response. Large dislocation densities and local hardening is observed when macroscopic strains vary over microns. A similar effect is expected in a cracked solid as the conditions ahead of a flaw resemble those at the tip of an indenter, leading to significantly higher crack tip stresses relative to conventional plasticity predictions. However, validating experimentally quantitative features specific to the various gradient plasticity models proposed is a challenging task. A direct comparison is possible in terms of crack tip opening displacements – less crack tip blunting is predicted using gradient-enriched constitutive models, and this seems to show a better agreement with experiments (Martínez-Pañeda *et al*., 2016b). In addition, the plastic stress intensity factors presented open the door for establishing indirect correlations with fracture toughness measurements.

# 7. Conclusions

We use CMSG plasticity, a first-order strain gradient plasticity formulation based on the Taylor dislocation model, to gain insight into plastic strain gradient effects on crack tip mechanics. The main conclusions of our combined analytical and numerical study are the following:

1. Plastic strain gradients elevate crack tip stresses relative to conventional plasticity solutions. A parametric study is conducted to map the regions of gradient plasticity significance.

2. The angular distribution of hoop and effective stresses in CMSG plasticity shows sensitivity to the distance ahead of the crack, implying that crack tip fields in CMSG plasticity do not have a separable form of solution.

3. The numerical analysis shows that, unlike the higher-order MSG plasticity model, the power of stress singularity in the lower-order CMSG plasticity theory is a function of the plastic work hardening exponent $N$.

4. Analytical and numerical approaches are employed to formulate novel amplitude and plastic stress intensity factors for strain gradient plasticity. A generalized $J$-integral for strain gradient plasticity is derived and used to characterize the nonlinear factors

5. The newly derived plastic stress intensity factor is uniform within the range $r/l = 0.001$ to $0.1$ and, as in conventional plasticity, exhibits a sensitivity to the strain hardening exponent $N$. However, the opposite trend is observed relative to the classic HRR solution as in the power of the stress singularity. Also, the sensitivity of the proposed fracture resistance parameter $K_p$ to the coupled effects of material length scale $l$ and plastic work hardening $N$ is established.

Potential avenues for future work include the coupling with damage mechanics-like models, to enable gradient-enriched damage assessment, and the use of the newly developed plastic stress intensity factors in conjunction with probabilistic Weibull-type schemes to determine the probability of failure of structural materials susceptible to cleavage fracture.

**Acknowledgments**

The authors ([1]VS, AT, AnT) gratefully acknowledge the financial support of the Russian Science Foundation under the Project 20-19-00158.

## Appendix 1. Derivation of the strain energy density

The elastic strain energy density in the expression (33) is described by the following equation:

$$\left(\frac{dW}{dV}\right)_E = \frac{1}{2}\sigma_{ij}\varepsilon_{ij} = \frac{1}{2}\left(\sigma_{xx}\varepsilon_{xx} + \sigma_{yy}\varepsilon_{yy} + \sigma_{zz}\varepsilon_{zz} + \sigma_{xy}\gamma_{xy} + \sigma_{yz}\gamma_{yz} + \sigma_{zx}\gamma_{zx}\right) =$$
$$= \frac{1}{2}\left[\sigma_{xx}\frac{\partial u}{\partial x} + \sigma_{yy}\frac{\partial v}{\partial y} + \sigma_{zz}\frac{\partial w}{\partial z} + \sigma_{xy}\left(\frac{\partial v}{\partial x} + \frac{\partial u}{\partial y}\right) + \sigma_{yz}\left(\frac{\partial w}{\partial y} + \frac{\partial v}{\partial z}\right) + \sigma_{zx}\left(\frac{\partial u}{\partial z} + \frac{\partial w}{\partial x}\right)\right]$$ (A1)

A series expansion may be performed to express the stresses and displacements in terms of the local polar coordinates $(r, \theta)$. Consider the first two terms proportional to $r^{-1/2}$, and use the following stress and displacement expansion in the form of equations

$$\sigma_{xx} = \frac{K_1}{\sqrt{2\pi r}}\cos\frac{\theta}{2}\left(1 - \sin\frac{\theta}{2}\sin\frac{3\theta}{2}\right) - \frac{K_2}{\sqrt{2\pi r}}\sin\frac{\theta}{2}\left(2 + \cos\frac{\theta}{2}\cos\frac{3\theta}{2}\right) + T$$

$$\sigma_{yy} = \frac{K_1}{\sqrt{2\pi r}}\cos\frac{\theta}{2}\left(1 + \sin\frac{\theta}{2}\sin\frac{3\theta}{2}\right) + \frac{K_2}{\sqrt{2\pi r}}\sin\frac{\theta}{2}\cos\frac{\theta}{2}\cos\frac{3\theta}{2}$$ (A2)

$$\sigma_{xy} = \frac{K_1}{\sqrt{2\pi r}}\sin\frac{\theta}{2}\cos\frac{\theta}{2}\cos\frac{3\theta}{2} + \frac{K_2}{\sqrt{2\pi r}}\cos\frac{\theta}{2}\left(1 - \sin\frac{\theta}{2}\sin\frac{3\theta}{2}\right)$$

$$u = \frac{K_1}{G}\sqrt{\frac{r}{2\pi}}\cos\frac{\theta}{2}\left[\frac{1}{2}(\kappa - 1) + \sin^2\left(\frac{\theta}{2}\right)\right] + \frac{K_2}{G}\sqrt{\frac{r}{2\pi}}\sin\frac{\theta}{2}\left[\frac{1}{2}(\kappa + 1) + \cos^2\left(\frac{\theta}{2}\right)\right] +$$
$$+ \frac{(1-\eta)\sigma}{8G}\{r[\cos(\theta + 2\alpha) + \kappa\cos(\theta - 2\alpha) - 2\sin\theta\sin 2\alpha]\} + \frac{T \cdot l}{8G}(\kappa + 1)$$ (A3)

$$v = \frac{K_1}{G}\sqrt{\frac{r}{2\pi}}\sin\frac{\theta}{2}\left[\frac{1}{2}(\kappa + 1) - \cos^2\left(\frac{\theta}{2}\right)\right] + \frac{K_2}{G}\sqrt{\frac{r}{2\pi}}\cos\frac{\theta}{2}\left[\frac{1}{2}(1 - \kappa) + \sin^2\left(\frac{\theta}{2}\right)\right] +$$
$$+ \frac{(1-\eta)\sigma}{8G}\{r[\sin(2\alpha - \theta) + \kappa\sin(2\alpha + \theta) - 2\sin\theta\cos 2\alpha]\} + \frac{T \cdot l}{8G}\tan 2\alpha(\kappa + 1)$$

$\kappa = 3 - 4\nu$ - plane strain, $\kappa = (3-\nu)/(1+\nu)$ - plane stress.

To apply Eq.(12) for the strain energy density parameter calculation it is necessary to obtain both the strain and displacement components. The relations between the components of the displacements in different coordinate systems are given by equations

$$\frac{\partial u_i}{\partial x} = \frac{\partial u_i}{\partial r}\cos\theta - \frac{\partial u_i}{\partial \theta}\cdot\frac{\sin\theta}{r}, \quad \frac{\partial u_i}{\partial y} = \frac{\partial u_i}{\partial r}\sin\theta + \frac{\partial u_i}{\partial \theta}\cdot\frac{\cos\theta}{r},$$ (A4)

while the relations between the strain and displacement components are described with the well-known formulae by Cauchy

$$\varepsilon_{xx} = \frac{\partial u}{\partial x}, \varepsilon_{yy} = \frac{\partial v}{\partial y}, \gamma_{xy} = \frac{\partial u}{\partial y} + \frac{\partial v}{\partial x}$$ (A5)

Applying Eqs. (13) and (14) to (12) and (15) to (16) and performing the necessary algebra, the expansion of the strain energy density field results in the dimensionless form

$$\overline{W}_E = \overline{K}_1^2 \frac{(1+\nu)}{2\pi} \overline{r}^{(-1)} a_{11} \tag{A6}$$

where

$$a_{11} = F_{x1}e_{x1} + F_{y1}e_{y1} + F_{xy1}(g_{11} + g_{21}) \tag{A7}$$

$$F_{x1} = \cos\frac{\theta}{2}\left(1 - \sin\frac{\theta}{2}\sin\frac{3\theta}{2}\right), \quad F_{y1} = \cos\frac{\theta}{2}\left(1 + \sin\frac{\theta}{2}\sin\frac{3\theta}{2}\right),$$

$$F_{xy1} = \cos\frac{\theta}{2}\sin\frac{\theta}{2}\cos\frac{3\theta}{2}$$

$$e_{x1} = \frac{1}{2}\left[\frac{1}{2}(\kappa - 1) + \sin^2\frac{\theta}{2}\right]\cos\frac{\theta}{2} - \sin\frac{\theta}{2}\sin\theta\cos^2\frac{\theta}{2}$$

$$e_{y1} = \frac{1}{2}\left[\frac{1}{2}(\kappa + 1) - \cos^2\frac{\theta}{2}\right]\cos\frac{\theta}{2} + \cos\frac{\theta}{2}\cos\theta\sin^2\frac{\theta}{2}$$

$$g_{11} = \frac{1}{2}\left[\frac{1}{2}(\kappa + 1) - \cos^2\frac{\theta}{2}\right]\left(-\sin\frac{\theta}{2}\right) - \cos\frac{\theta}{2}\sin\theta\sin^2\frac{\theta}{2}$$

$$g_{21} = \frac{1}{2}\left[(\kappa - 1) + \sin^2\frac{\theta}{2}\right]\left(-\sin\frac{\theta}{2}\right) + \sin\frac{\theta}{2}\cos\theta\cos^2\frac{\theta}{2}$$

$$I_{el} = \int_{-\pi}^{+\pi} a_{11}\cos\theta \, d\theta = -0.726493 \tag{A8}$$